\documentclass[fleqn,usenatbib]{mnras}

\usepackage{newtxtext,newtxmath}

\usepackage[T1]{fontenc}
\usepackage{ae,aecompl}

\usepackage{graphicx}	
\graphicspath{{./figures/}}

\usepackage{amsmath}	
\usepackage{amssymb}	

\usepackage{units}

\usepackage[usenames,dvipsnames]{color}

\usepackage{multicol}

\newcommand{\nuclei}[2]{\ensuremath{\mathrm{^{#1}#2}}}

\newcommand{\oxygen}[1][16]{\nuclei{#1}{O}}
\newcommand{\fluorine}[1][19]{\nuclei{#1}{F}}
\newcommand{\neon}[1][20]{\nuclei{#1}{Ne}}
\newcommand{\sodium}[1][23]{\nuclei{#1}{Na}}
\newcommand{\magnesium}[1][24]{\nuclei{#1}{Mg}}

\newcommand{\code}[1]{\textsc{#1}}
\newcommand{\mesa}{\code{MESA}}
\newcommand{\MESA}{\mesa}

\newcommand{\vt}{\texttt{varcontrol\_target}}
\newcommand{\mdc}{\texttt{mesh\_delta\_coeff}}

\newcommand{\kB}{\ensuremath{k_\mathrm{B}}} 
\newcommand{\mb}{\ensuremath{m_\mathrm{u}}} 
\newcommand{\me}{\ensuremath{m_\mathrm{e}}} 

\newcommand{\Tc}{\ensuremath{T_{\mathrm{\!c}}}} 

\newcommand{\rhoc}{\ensuremath{\rho_{\mathrm{c}}}} 
\newcommand{\gcc}{\ensuremath{\mathrm{g\,cm^{-3}}}} 

\newcommand{\Ye}{\ensuremath{Y_{\mathrm{e}}}} 

\newcommand{\Msun}{\ensuremath{\mathrm{M}_{\sun}}} 
\newcommand{\Mdot}{\ensuremath{\dot{M}}} 
\newcommand{\Msunyr}{\ensuremath{\rm \Msun\,yr^{-1}}} 

\newcommand{\EF}{\ensuremath{E_\mathrm{F}}} 
\newcommand{\Qg}{\ensuremath{Q_\mathrm{g}}} 

\newcommand{\Xu}{\ensuremath{X_{\mathrm{u}}}} 

\newcommand{\cv}{\ensuremath{c_{\mathrm{v}}}} 

\newcommand{\logT}{\ensuremath{\log(T/\mathrm{K})}} 
\newcommand{\logTc}{\ensuremath{\log(T_{\rm c}/\mathrm{K})}} 
\newcommand{\logRho}{\ensuremath{\log(\rho/\gcc)}} 
\newcommand{\logRhoc}{\ensuremath{\log(\rhoc/\gcc)}} 

\newcommand{\Jpi}[1]{\ensuremath{J_{\mathrm{#1}}^{\pi}}}
\newcommand{\Ei}{\ensuremath{E_\mathrm{i}}} 
\newcommand{\Ef}{\ensuremath{E_\mathrm{f}}} 

\newcommand{\Ee}{\ensuremath{E_\mathrm{e}}} 

\newcommand{\gradad}{\ensuremath{\nabla_{\rm ad}}} 
\newcommand{\gradT}{\ensuremath{\nabla_T}} 

\newcommand{\logft}{\ensuremath{\log(ft/{\rm s})}} 

\newcommand{\revision}{Revision 0d251e4b492c0eea72b89b77c7a7c43077b13e27}

\newcommand{\SQB}{SQB15}
\newcommand{\tcross}{t_\textrm{cross}}
\newcommand{\tcool}{t_\textrm{cool}}

\title[Urca-process Cooling in ONe White Dwarfs]{The Importance of Urca-process Cooling in Accreting ONe White Dwarfs}
\author[Schwab et al.]{
Josiah Schwab$^{1,2}$\thanks{Hubble Fellow; E-mail: jwschwab@ucsc.edu},
Lars Bildsten$^{3}$,
Eliot Quataert$^{1}$
\\
$^1${Physics Department and Astronomy Department and Theoretical Astrophysics Center, University of California, Berkeley, CA 94720, USA} \\
$^2${Department of Astronomy and Astrophysics, University of California, Santa Cruz, CA 95064, USA} \\
$^3${Kavli Institute for Theoretical Physics and Department of Physics, University of California, Santa Barbara, CA 93106, USA} \\
}

\date{\revision}

\pubyear{2017}

\begin{document}
\label{firstpage}
\pagerange{\pageref{firstpage}--\pageref{lastpage}}
\maketitle

\begin{abstract}
  We study the evolution of accreting oxygen-neon (ONe) white dwarfs
  (WDs), with a particular emphasis on the effects of the presence of
  the carbon-burning products \sodium[23] and \magnesium[25].  These
  isotopes lead to substantial cooling of the WD via the
  \magnesium[25]-\sodium[25], \sodium[23]-\neon[23], and
  \sodium[25]-\neon[25] Urca pairs.  We derive an analytic formula for
  the peak Urca-process cooling rate and use it to obtain a simple
  expression for the temperature to which the Urca process cools the
  WD. Our estimates are equally applicable to accreting carbon-oxygen
  WDs.  We use the Modules for Experiments in Stellar Astrophysics
  (\MESA) stellar evolution code to evolve a suite of models that
  confirm these analytic results and demonstrate that Urca-process
  cooling substantially modifies the thermal evolution of accreting
  ONe WDs.  Most importantly, we show that \MESA\ models with lower
  temperatures at the onset of the \magnesium[24] and \sodium[24]
  electron captures develop convectively unstable regions, even when
  using the Ledoux criterion.  We discuss the difficulties that we
  encounter in modeling these convective regions and outline the
  potential effects of this convection on the subsequent WD evolution.
  For models in which we do not allow convection to operate, we find
  that oxygen ignites around a density of $\logRhoc \approx 9.95$,
  very similar to the value without Urca cooling.  Nonetheless, the
  inclusion of the effects of Urca-process cooling is an important
  step in producing progenitor models with more realistic temperature
  and composition profiles which are needed for the evolution of the
  subsequent oxygen deflagration and hence for studies of the
  signature of accretion-induced collapse.
\end{abstract}



\section{Introduction}

In the Urca process, first discussed by \citet{Gamow1941}, repeated
electron-capture and beta-decay reactions give rise to neutrino
emission.  When this occurs in a stellar interior where the
neutrinos are able to free-stream out of the star---such as in a white
dwarf (WD)---it becomes an active cooling process.
\citet{Tsuruta1970} calculated analytic approximations to the energy
loss rates from the Urca process and compiled a list of 132 pairs of
isotopes that contribute to these energy losses.  \citet{Paczynski1973}
applied these results in a study of the temperature evolution of
degenerate carbon-oxygen (CO) cores, demonstrating that this cooling
can shift the density at which pycnonuclear carbon ignition occurs.

In \citet{Paczynski1973} the odd mass number nuclei that participate in
the Urca process were assumed to have cosmic abundances.  Carbon
burning, however, produces significant mass fractions of $\sodium[23]$
and $\magnesium[25]$. Therefore Urca-process cooling will be significantly
more important in stars with oxygen-neon (ONe) compositions, where the
material has already been processed by carbon burning \citep{Iben1978},
such as in the cores of super-asymptotic giant branch stars
\citep{Toki2013, Jones2013}.

In \citet{Schwab2015}, hereafter referred to as \SQB, we developed an
analytic and numerical understanding of the evolution of ONe WDs
towards accretion-induced collapse (AIC) in which we considered only
\magnesium[24], \neon[20], and \oxygen[16].  In this work, we extend
and modify this understanding to include additional odd mass number
isotopes generated during carbon-burning, namely \sodium[23] and
\magnesium[25].  We demonstrate analytically and numerically that
significant temperature changes occur due to Urca-process cooling and
we illustrate its effect on the subsequent evolution.  Most
importantly, we find that the Urca process alters the temperature
profile of the WD in such a way that regions of the WD become
convectively unstable after the electron captures on \magnesium[24]
occur.

In Section~\ref{sec:urca}, we provide an overview of the microphysics
of the Urca process and identify the important isotopes and their
threshold densities.  In Section~\ref{sec:analytics}, we make analytic
estimates of the importance of Urca-process cooling in accreting ONe
WDs. In Section~\ref{sec:mesa-calculations}, we discuss how we use the
\MESA\ stellar evolution code to demonstrate the effects of
Urca-process cooling.  In Section~\ref{sec:forbidden}, we discuss and
characterize the effects of incomplete nuclear data.  In
Section~\ref{sec:onset-electr-capt-24}, we demonstrate and explain the
onset of convective instability in our \MESA\ models.  In
Section~\ref{sec:subsequent-evolution}, we discuss the evolution of the
WD up to oxygen ignition. In Section~\ref{sec:conclusions}, we conclude.

\section{The Urca Process}
\label{sec:urca}

Take two nuclei $a \equiv (Z,A)$ and $b \equiv (Z-1, A)$ that are
connected by an electron-capture transition
\begin{equation}
  \label{eq:electron-capture}
  (Z,A) + e^- \to (Z-1,A) + \nu_e
\end{equation}
and beta-decay transition
\begin{equation}
  \label{eq:beta-decay}
  (Z-1,A) \to (Z,A) + e^- + \bar{\nu}_e
\end{equation}
where $Z$ and $A$ are respectively the atomic number and mass number
of the nucleus.  In all of the electron-capture transitions considered
here, there is a threshold energy required for the electron.  In a
cold, degenerate plasma, electrons with sufficient energy will become
available when the Fermi energy $\EF$ is equal to the energy
difference between the parent and daughter states $Q_0$, which
includes both the nuclear rest mass and the energy associated with
excited states.  In the limit of relativistic electrons, this corresponds to a threshold density
\begin{equation}
  \rho_{0} \approx \unit[1.8 \times 10^{9}]{\gcc}\, \left(\frac{\Ye}{0.5}\right)^{-1} \left(\frac{|Q_0|}{\unit[5]{MeV}}\right)^{3} ~.
  \label{eq:threshold}
\end{equation}
where $\Ye$ is the electron fraction.

\subsection{Cooling Rate}

At the threshold density the rates of electron capture and beta decay are
comparable.  Since each reaction produces a neutrino that free-streams
out of the star, this is a cooling process.

Suppose the total number density of the two isotopes in the Urca pair
is $n_u = n_a + n_b$.  Because the time-scales for electron capture and
beta decay are short compared to the evolutionary time-scale of the
system, an equilibrium is achieved.  The relative abundances are then
given by the detailed balance condition
$n_a \lambda_\mathrm{ec} + n_b \lambda_\beta = 0$.  Under this
assumption, the specific neutrino cooling rate from the Urca process
can be written as
\begin{equation}
  \label{eq:urca-specific}
  \epsilon_\mathrm{u} = \frac{n_u}{\rho} C= \frac{X_u}{A_u \mb} C
\end{equation}
where $X_u$ is the mass fraction of the Urca pair, $A_u$ is its atomic
mass number, $\mb$ is the atomic mass unit, and
\begin{equation}
  \label{eq:urca-C}
  C = \frac{\varepsilon_{\nu, \mathrm{ec}} \lambda_\beta + \varepsilon_{\nu, \beta} \lambda_\mathrm{ec}}{\lambda_\beta + \lambda_\mathrm{ec}}~.
\end{equation}

In Appendix~\ref{sec:urca-physics}, we write out the full expressions
for the rates ($\lambda$) and neutrino loss rates ($\varepsilon_\nu$)
for electron capture and beta decay necessary to evaluate
equation~\eqref{eq:urca-C}.  The key result is that the Urca-process cooling
rate for an allowed ground state to ground state transition is sharply
peaked at \mbox{$\EF = |\Qg|$} and that the maximum value of $C$ is
\begin{equation}
  \label{eq:urca-C-max}
  C_\mathrm{max} = \frac{ 7 \upi ^4 \ln 2}{60}\frac{ \me c^2}{(ft)_{\beta} + (ft)_{\mathrm{ec}}} \left(\frac{\kB T}{\me c^2}\right)^4 \left(\frac{\Qg}{\me c^2}\right)^2 \exp({\upi\alpha Z})~,
\end{equation}
where $ft$ is the comparative half-life, $\alpha$ is the fine structure constant, and $\Qg$ is the threshold energy for the ground state to ground state transition.

\subsection{Isotopes and Transitions}
\label{sec:isos}

\begin{table*}
  \centering
  \caption{A summary of the key weak reactions that occur in accreting
    ONe WDs.  Only the lowest energy allowed transition from the
    ground state is listed, since this typically sets the threshold
    density; this is not an exhaustive list of the transitions
    considered in this work.  Electron captures convert the initial
    isotope to the final isotope.  $\Qg$ is the rest mass difference
    between the ground states of the isotopes (in $\unit[]{MeV}$); unlike in the similar
    tabulation in \SQB\ we have already accounted for the electron
    rest mass.  $\Ei$ and $\Ef$ are the excitation energies of the
    initial and final states, relative to the ground state (in
    $\unit[]{MeV}$).  $\Jpi{i}$ and $\Jpi{f}$ are the spins and
    parities of the initial and final states.  $ft$ is the
    comparative half-life (in $\unit[]{s}$) for this transition.
    $Q_0$ is the threshold energy difference (in
    $\unit[]{MeV}$); this includes the energy associated with excited states. $\rho_0$ is the approximate density (in
    $\unit[]{g\,cm^{-3}}$) at which the reaction occurs (as defined in
    equation~\ref{eq:threshold}).  Effect indicates whether the net
    effect of the weak reactions is to cool the plasma via
    Urca-process cooling (odd mass number) or heat the plasma via
    exothermic electron captures (even mass number).  The nuclear data is drawn from the literature
\citep{Tilley1998,Firestone2007a,Firestone2007b,Firestone2009,MartinezPinedo2014}.}
  \label{tab:weak-data}
  \begin{tabular}{llrrrrrrrrrl}
    Initial        & Final         & \Qg    & \Ei   & \Jpi{i} & \Ef   & \Jpi{f} & $\log (ft)$ & $Q_0$  & $\log \rho_0$ & Effect & Notes  \\
    \hline
    \magnesium[25] & \sodium[25]   & -4.346 & 0.000 & $5/2^+$ & 0.000 & $5/2^+$ & 5.26        & -4.346 & 9.07          & Cool   &        \\
    \sodium[23]    & \neon[23]     & -4.887 & 0.000 & $3/2^+$ & 0.000 & $5/2^+$ & 5.27        & -4.887 & 9.22          & Cool   &        \\
    \magnesium[24] & \sodium[24]   & -6.026 & 0.000 & $0^+$   & 0.472 & $1^+$   & 4.82        & -6.498 & 9.60          & Heat   & \\
    \sodium[24]    & \neon[24]     & -2.978 & 0.000 & $4^+$   & 3.972 & $4^+$   & 6.21        & -6.950 & 9.69          & Heat   & $^{a}$ \\
    \sodium[25]    & \neon[25]     & -7.761 & 0.090 & $3/2^+$ & 0.000 & $1/2^+$ & 4.41        & -7.671 & 9.81          & Cool   & $^{b}$  \\
    \neon[20]      & \fluorine[20] & -7.536 & 0.000 & $0^+$   & 1.057 & $1^+$   & 4.38        & -8.593 & 9.96          & Heat   & $^{a,c}$ \\
    \neon[23]      & \fluorine[23] & -8.991 & 0.000 & $5/2^+$ & 0.000 & $5/2^+$ & 5.72        & -8.991 & 10.02         & Cool   & $^{d}$ \\
    \hline
    \multicolumn{12}{l}{$^{a}$ this reaction is affected by a nonunique second forbidden transition; see Section~\ref{sec:forbidden}} \\
    \multicolumn{12}{l}{$^{b}$ the ground state has $J^\pi = 5/2^+$; for the relevant temperatures this low-lying excited state is populated} \\
    \multicolumn{12}{l}{$^{c}$ the \fluorine[20] will immediately undergo an electron capture to form \oxygen[20]}                      \\
    \multicolumn{12}{l}{$^{d}$ the oxygen deflagration begins before our models reach this density}                                     \\
  \end{tabular}
\end{table*}


Using a nuclear reaction network with 244 species and analytic weak
reaction rates, \citet{Iben1978} identified Urca pairs for which the
neutrino loss rates rival or exceed thermal neutrino losses.  In
material processed by carbon burning, the two most abundant odd mass
number isotopes are \sodium[23] and \magnesium[25] and thus the most
important Urca pairs have $A=23$ and $25$ \citep[see figure 2
in][]{Iben1978}.  Therefore, we restrict our attention to these
isotopes, neglecting possible small contributions from $A=21$ and $29$
isotopes.

\begin{figure*}
  \centering
  \includegraphics[width=\columnwidth]{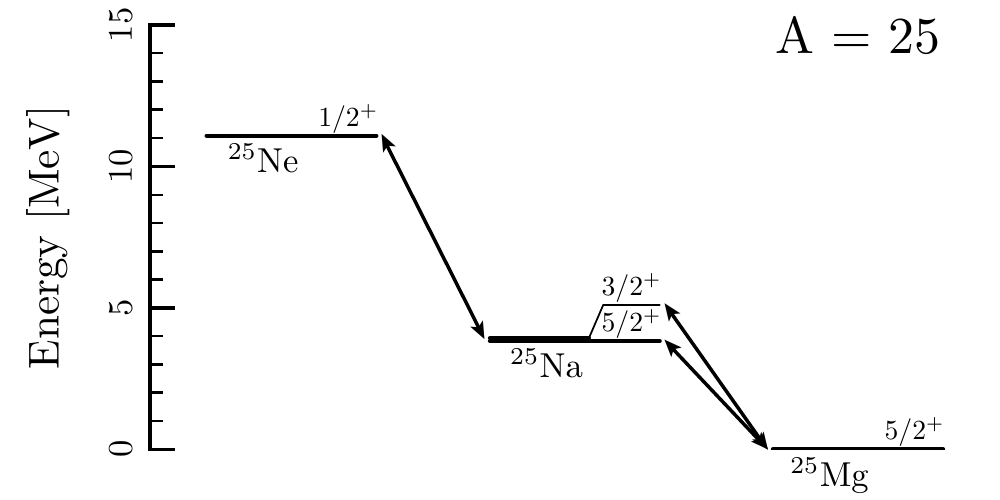}
  \includegraphics[width=\columnwidth]{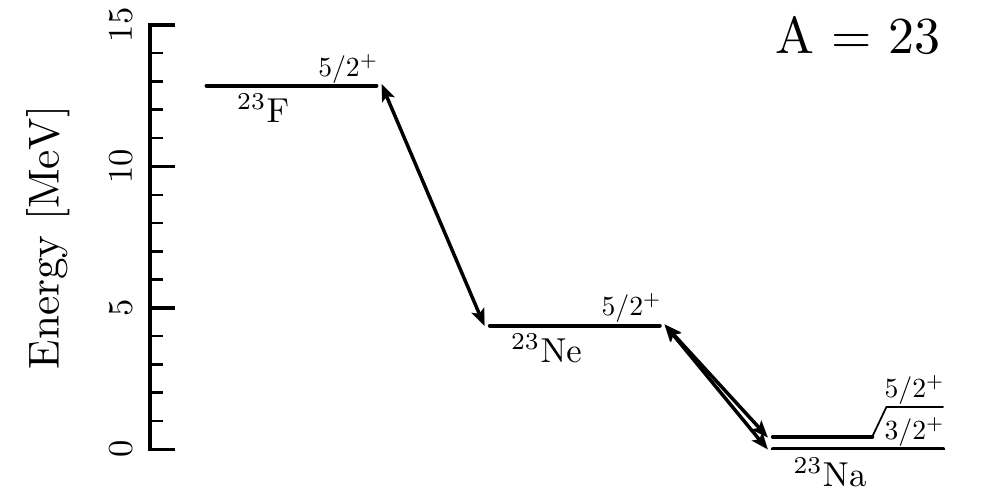}
  \caption{Energy level diagrams for the $A=25$ (left) and $A=23$
    (right) nuclei that we consider.  The $J^{\pi}$ value is indicated
    at the right of each level and is sometimes given an arbitrary
    offset (indicated via a thin line) in order to enhance
    legibility. The most important transitions we consider are
    indicated with arrows.  The transition between \neon[25] and
    \sodium[25] is to/from the low-lying first excited state of
    \sodium[25].}
  \label{fig:levels2523}
\end{figure*}

The nuclear data (energy levels and $ft$ values) required for this
calculation are drawn from the literature
\citep{Tilley1998,Firestone2007a,Firestone2007b,Firestone2009,MartinezPinedo2014}.
Table~\ref{tab:weak-data} summarizes this data.
Fig.~\ref{fig:levels2523} shows a simplified level structure
(excluding excited states $\ga 1\, \unit{MeV}$ above the ground state)
of the $A=23$ and $A=25$ nuclei that we consider.  In Section~\ref{sec:forbidden}, we will discuss the $A=20$ and $A=24$ nuclei in more detail.

While this work was in preparation, new weak reaction rate tables for
A=17-28 nuclei were published by \citet{Suzuki2016}.  In
Appendix~\ref{sec:suzuki-tables} we compare our fiducial model with a
\MESA\ calculation using those tables.  We find good agreement.

\section{Analytic Estimates}
\label{sec:analytics}

\begin{figure}
  \centering
  \includegraphics[width=\columnwidth]{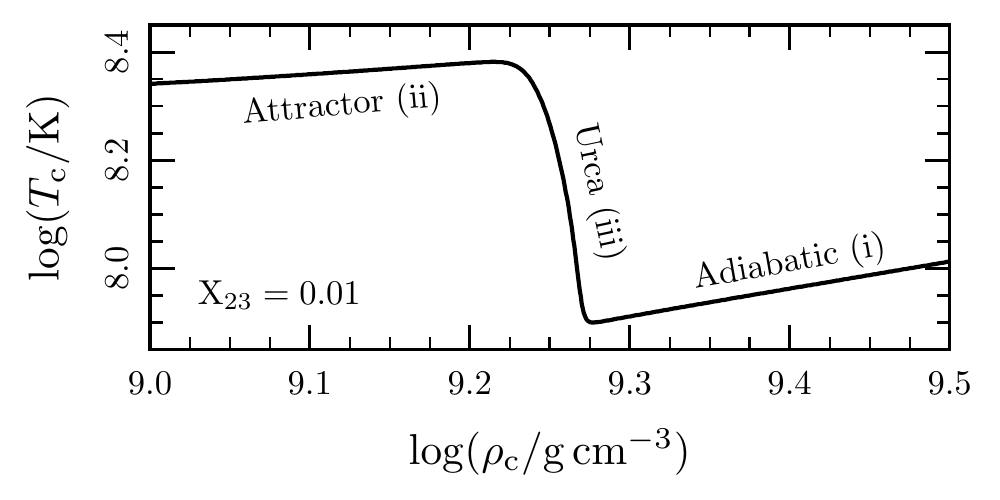}
  \caption{The schematic evolution of central density and temperature
    in a ONe WD accreting at $10^{-6} \Msunyr$.\@ The black line shows
    the result of evolving a \mesa\ model that has an initial mass
    fraction of 0.01 \sodium[23].  The three regimes discussed in
    Section~\ref{sec:analytics} are labeled.}
  \label{fig:schematic}
\end{figure}

The energy equation for material in a spherically symmetric star is
\begin{equation}
  \label{eq:energy-eqn}
  T \frac{ds}{dt} = \epsilon - \frac{\upartial L}{\upartial M}
\end{equation}
where $\epsilon$ is the specific energy generation rate, which
includes nuclear reactions, neutrino processes, etc.  For these
estimates, we will consider only the effects of neutrino losses, which
we sub-divide into $\epsilon_\nu$ (thermal neutrino loss rate) and
$\epsilon_{\mathrm{u}}$ (Urca process neutrino loss rate).  In the
centre of these rapidly accreting WDs, $\upartial L / \upartial M$ is negligible, and
therefore
\begin{equation}
  \label{eq:energy-eqn-center}
  -\left(\epsilon_\nu + \epsilon_{\mathrm{u}}\right) = \Tc\ \cv \left[\frac{d \ln \Tc}{dt} - (\Gamma_3 -1) \frac{d \ln \rhoc}{dt}\right]~,
\end{equation}
where $\cv$ is the specific heat at constant volume and
$\Gamma_3 -1 = \left(d\ln T/d\ln \rho \right)_{\mathrm{ad}}$.  Depending on
  which terms dominate, there are three regimes for the evolution of
  the central temperature:
\begin{enumerate}
\item When the left hand side of equation~\eqref{eq:energy-eqn-center}
  is negligible, the central temperature will evolve along an adiabat.
\item When $\epsilon_\nu$ dominates the left hand side of
  equation~\eqref{eq:energy-eqn-center}, the temperature will evolve
  towards (and then along) the attractor solution discussed by
  \citet{Paczynski1973}, \SQB, and \citet{Brooks2016},
  in which thermal neutrino cooling and compressional heating balance.
  Because of the neutrino
  losses, this attractor solution is shallower (in $T$-$\rho$ space)
  than an adiabat, though it still has positive slope.
\item When $\epsilon_u$ dominates the left hand side of
  equation~\eqref{eq:energy-eqn-center}, the temperature will
  decrease.  Because the Urca-process cooling is sharply peaked in
  $\EF$, this will occur at nearly fixed density.
\end{enumerate}
These three regimes are illustrated in Fig.~\ref{fig:schematic},
which shows the evolution of the central density and temperature in
one of our \MESA\ models, centered on the density where cooling due to
the \sodium[23]-\neon[23] Urca pair occurs.

It is useful to estimate the magnitude of the temperature decrease
caused by the Urca process (regime iii).  As we will show, this
depends primarily on the mass fraction of the Urca pair and the
rate at which the core is being compressed.  \citet{Paczynski1973}
provides a fitting formula for the temperature change, obtained though
careful numerical integration; however this result is unsuitable for
our purposes, as it assumes that the value of
$d\ln \rho/dt$ is that of a CO core growing via stable He-shell
burning, as set by the core mass-luminosity relation.

We assess the Urca-process cooling via a simpler argument.  As a
result of accretion, the core is being compressed on a time-scale
\begin{equation}
  \label{eq:trho}
  t_{\rho} = \left(\frac{d \ln \rho_c}{dt} \right)^{-1} = \left( \frac{d \ln \rho_c}{d \ln M} \right)^{-1} \frac{M}{\dot{M}}~~~.
\end{equation}
For an ideal, zero-temperature white dwarf, in the range
$9 \la \logRhoc \la 10$ and with $\Ye \approx 0.5$,
\SQB\ give the approximate result that
\begin{equation}
  \label{eq:trho-numerical}
  t_{\rho} \approx \unit[5\times10^4]{yr}\,\rho_9 ^{-0.55} \dot{M}_{-6}^{-1}
\end{equation}
where $\rho_9 = \rho / (\unit[10^9]{g~cm^{-3}})$ and
$\dot{M}_{-6} = \dot{M}/ ({\unit[10^{-6}]{\Msunyr}})$.

The cooling from an individual Urca pair peaks when $\EF = \Qg$, and
is significant for only $\Delta \EF \approx 3 \kB T$ centered around this
peak (see Appendix~\ref{sec:urca-physics}, in particular equation~\ref{eq:c-taylor}).  We can estimate the width of the peak (in density) as
$\Delta \ln \rho \approx 3 \Delta \EF / \EF$.  Therefore, the time-scale
for a parcel to cross the cooling region is
\begin{align}
  \label{eq:tcross}
  \tcross \approx \left(\frac{9 \kB T}{\EF}\right) t_\rho & \approx 2 \times 10^{-2}\, T_8 \rho_9^{-1/3} t_\rho \\
& \approx \unit[1\times10^{3}]{yr}\, T_8 \rho_9^{-0.88} \dot{M}_{-6}^{-1} ~.
\end{align}



At the density where the Urca-process cooling peaks, the cooling time-scale
$\tcool$ is
\begin{equation}
  \label{eq:tcool}
  \tcool = \frac{c_{\mathrm{v}} T}{\epsilon_{\mathrm{max}}} = \frac{3 \kB T A_{\mathrm{u}}}{\bar{A} \Xu C_{\mathrm{max}}}
\end{equation}
where we have taken $\epsilon_{\mathrm{max}}$ from the combination of
equations~\eqref{eq:urca-specific} and \eqref{eq:urca-C-max}, and we
have assumed the specific heat is that given by the Dulong-Petit law
($c_{\mathrm{v}} = 3 \kB / \bar{A}$).  Assuming
$A_{\mathrm{u}} \approx \bar{A}$,
\begin{equation}
  \label{eq:tcool-num}
  \tcool \approx \unit[4\times10^2]{yr}\, T_8^{-3}
 \left(\frac{\Xu}{0.01}\right)^{-1}
 \left(\frac{Q_{\mathrm{g}}}{\unit[5]{MeV}}\right)^{-2}
 \left(\frac{ft}{\unit[10^5]{s}}\right) ~.
\end{equation}

When the core reaches a density where Urca-process cooling will begin, its
initial temperature will have been set by its evolution in regimes (i)
or (ii).  If $\tcross > \tcool$ initially, since $\tcool$ increases
more rapidly with decreasing temperature than $\tcross$, the core will evolve
towards the condition $\tcross \approx \tcool$.  When this condition
is reached, the Urca-process cooling will effectively shut off, since the core
will evolve out of the cooling region before significant additional
cooling occurs.  If $\tcross < \tcool$ initially---which is never true
in the cases we consider---then significant Urca-process cooling will
not occur.

\begin{figure}
  \centering
  \includegraphics[width=\columnwidth]{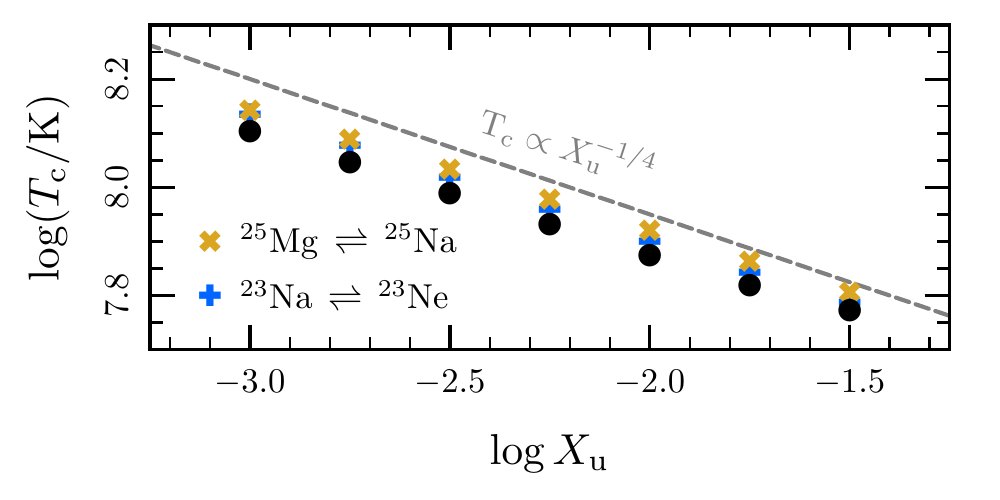}
  \caption{The minimum central temperature ($\Tc$) reached after
    Urca-process cooling as a function of the mass fraction in
    the Urca pair ($\Xu$).  The crosses (Xs) show models with an
    initial mass fraction $\Xu$ of \sodium[23] (\magnesium[25]). The
    solid black circles show models with initial mass fractions $\Xu$ of both
    \sodium[23] and \magnesium[25].  The dashed line shows the
    analytically expected scaling of equation~\eqref{eq:t-urca}.  }
  \label{fig:scalings-logx}
\end{figure}

\begin{figure}
  \centering
  \includegraphics[width=\columnwidth]{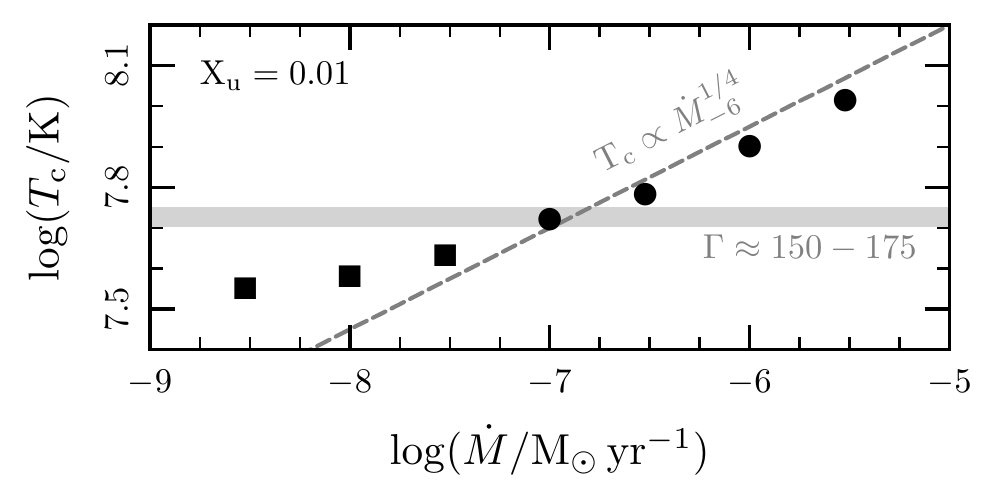}
  \caption{The minimum central temperature ($\Tc$) reached after Urca
    cooling as a function of the accretion rate ($\dot{M}$).  All
    models have $\Xu = 0.01$ as \sodium[23].  The solid grey band
    shows range of $\Gamma$ over which the latent heat of
    crystallization in released in \MESA.  Models that fully
    crystallized are marked with squares.  The dashed line shows the
    analytically expected scaling of equation~\eqref{eq:t-urca}.  }
  \label{fig:scalings-mdot}
\end{figure}

Therefore, the relation $\tcross \approx \tcool$ gives us an estimate
for the temperature to which each Urca pair will cool the star.
Combining equations~\eqref{eq:tcross} and \eqref{eq:tcool} and taking
the fiducial values $\Qg = \unit[5]{MeV}$, $ft = \unit[10^5]{s}$,
and using a density equal to the threshold density (equation
\ref{eq:threshold}) for this $\Qg$, we find that the temperature to
which the core cools is
\begin{equation}
  \label{eq:t-urca}
  T_{\mathrm{u}} \approx \unit[9 \times 10^7]{K}\, \dot{M}_{-6}^{1/4}  \left(\frac{\Xu}{0.01}\right)^{-1/4} ~.
\end{equation}
In order to validate this relation we ran a suite of \mesa\ models
varying $\Xu$ and $\dot{M}$.  These numerical results are shown in
Figs.~\ref{fig:scalings-logx} and \ref{fig:scalings-mdot} and are in
excellent agreement with the analytic scaling given in
equation~\eqref{eq:t-urca}.\footnote{The temperatures in the \MESA\
  models are $\approx 10$ per cent lower than this estimate,
  suggesting that a prefactor of $\unit[8\times10^7]{K}$ in
  equation~\eqref{eq:t-urca} would yield a slightly more accurate
  estimate.}  We will discuss the implications of this cooling on the
subsequent evolution in Section~\ref{sec:onset-electr-capt-24}.

We note that the Urca-process cooling can be significant enough to
cause the white dwarf to begin to crystallize.  At densities near the
\sodium[23] threshold density, the condition for this phase transition
$(\Gamma \approx 175)$ occurs at
$T \approx \unit[5 \times 10^{7}]{K}$.  Therefore, we expect
crystallization to begin when $\dot{M}_{-6} \Xu^{-1} \la \, 10$, at
which point the Urca-process cooling will begin extracting the latent heat
associated with the phase transition.  As shown in
Fig.~\ref{fig:scalings-mdot}, some of our models reach this regime.
However, we choose not to explore the interaction of crystallization and these weak reactions further.  Subsequent adiabatic compression and exothermic electron captures will cause the WD to be in the liquid state at the later times of primary interest.

\section{Details of \mesa\ Calculations}
\label{sec:mesa-calculations}

The calculations performed in this paper use \mesa\ version 9793
(released 2017-05-31).  As required by the \mesa\ manifesto, the
inlists necessary to reproduce our calculations will be posted on
\url{http://mesastar.org}.

\subsection{Initial Models}

We generate our initial models in the same manner as \SQB, except that
we stop relaxing the models at lower density
($\logRhoc = 8.6$) so that the Urca processes of
interest have not yet occurred.

Our models are initially chemically homogeneous.  The models shown as part of
the scaling studies in Section~\ref{sec:analytics} all have the
indicated abundances of \sodium[23] and \magnesium[25], a mass
fraction of 0.5 \oxygen[16], with the remainder as \neon[20].  In
Section~\ref{sec:subsequent-evolution}, we show results from four
compositions, identified as follows: \SQB, the composition used in
\SQB; this paper, a similar composition plus representative
mass fractions of \sodium[23] and \magnesium[25]; T13, a composition
based on the intermediate-mass star models of \citet{Takahashi2013};
and F15, a composition based on the intermediate-mass star models of
\citet{Farmer2015}.  The mass fractions of the isotopes present in
each named model are shown in Table~\ref{tab:compositions}.

\begin{table}
  \centering
  \caption{The set of compositions used in our \MESA\ models.  Each
    composition is referenced in the text by the identifier listed in
    the top row.  Each column lists the mass fractions of the isotopes
    (listed at left) that were included.  Dashes indicate that a
    particular isotope was not included.  The compositions T13 and F15
    are based on the intermediate-mass star models of
    \citet{Takahashi2013} and \citet{Farmer2015} respectively.}
  \label{tab:compositions}
  \begin{tabular}{rcccc}
    \hline
    Isotope & \SQB & This Paper & T13 & F15  \\
    \hline
    \oxygen[16]    & 0.500 & 0.500 & 0.480 & 0.490 \\
    \neon[20]      & 0.450 & 0.390 & 0.420 & 0.400 \\
    \neon[22]      & ---   & ---   & ---   & 0.018 \\
    \sodium[23]    & ---   & 0.050 & 0.035 & 0.060 \\
    \magnesium[24] & 0.050 & 0.050 & 0.050 & 0.030 \\
    \magnesium[25] & ---   & 0.010 & 0.015 & 0.002 \\
    \hline
  \end{tabular}
\end{table}

\subsection{Important \mesa\ Options}
\label{sec:mesa-options}

While our full inlists will be made publicly available, we highlight
some of the most important \mesa\ options used in the calculations.
This section assumes the reader is familiar with specific \mesa\
options.  Please consult the instrument papers \citep{Paxton2011,
  Paxton2013, Paxton2015} and the \mesa\
website\footnote{\url{http://mesa.sourceforge.net}} for a full
explanation of the meaning of these options.

Most importantly, we use the capability of \MESA\ to calculate weak
rates from input nuclear data developed in \SQB\ and validated in
\citet{Paxton2015, Paxton2016}.  The dangers of using coarse
tabulations of the relevant weak reaction rates has been emphasized by
\citet{Toki2013}; this choice circumvents these issues.  We activate these capabilities using
the options:
\begin{verbatim}
    use_special_weak_rates = .true.
    ion_coulomb_corrections = 'PCR2009'
    electron_coulomb_corrections = 'Itoh2002'
\end{verbatim}
Table~\ref{tab:weak-data} summarizes the weak reactions that we
include using this capability.  The files containing the input nuclear
data will be made available along with our inlists.

The \mesa\ equation of state \citep[][figure 1]{Paxton2011} contains a
transition from HELM \citep{Timmes2000b} to PC \citep{Potekhin2010}.
We set the location of this blend via the options
\begin{verbatim}
  log_Gamma_all_HELM = 0.60206d0 ! Gamma = 4
  log_Gamma_all_PC = 0.90309d0 ! Gamma = 8
\end{verbatim}
which ensures that the core of the WD is always treated using the PC
equation of state.  Rapid and significant composition changes will
occur as the weak equilibrium shifts.  Therefore, it is necessary to
ensure that all isotopes are included in the PC
calculation\footnote{The \mesa\ default is to only include isotopes
  with a mass fraction greater than 0.01 in the the PC equation of
  state calculation.  As the chemical composition changes, abundances
  rise above or fall below this threshold.  The sudden inclusion or
  exclusion of an isotope gives rise to a discontinuity in the
  equation of state.  While the jumps in the computed thermodynamic
  properties are small, the discontinuous nature of the changes leads
  to convergence problems in the Newton-Raphson solver as \MESA\
  iterates to find the next model.} by using the options:
\begin{verbatim}
    set_eos_PC_parameters = .true.
    mass_fraction_limit_for_PC = 0d0
\end{verbatim}

It is essential that we choose a temporal and spatial resolution that
will resolve the effects of Urca-process cooling and the exothermic
electron captures.  We discuss the the details of our approach in
Appendix~\ref{sec:convergence} and demonstrate that it leads to a
converged result.

The choice of convective criterion is important.  We use the Ledoux
criterion, which accounts for the effect of composition gradients on
the buoyancy.  The exothermic electron captures create temperature
gradients that would be unstable by the Schwarzschild criterion, but
are stabilized by the $\Ye$ gradients \citep{Miyaji1987}.
Convectively stable regions with such gradients are subject to
doubly-diffusive instabilities, but following the arguments in \SQB\
that suggest these regions will not have time to mix, we neglect the
effects of semiconvection.  These choices correspond to the \MESA\
options:
\begin{verbatim}
    use_Ledoux_criterion = .true.
    alpha_semiconvection = 0.0
\end{verbatim}

In Section~\ref{sec:onset-electr-capt-24} we will demonstrate that
convective instability can set in even when using the Ledoux
criterion.  Modeling this phase with standard MLT in \MESA\ proves
problematic and therefore most of the models shown use
\verb+mlt_option = 'none'+.  This choice means that convectively
unstable regions have the radiative temperature gradient and do not
experience any convective mixing.  A few of our models use a milder
restriction, preventing convection from modifying the temperature
gradient, but allowing for convective mixing.  This is achieved using
the control \verb+mlt_gradT_fraction = 0+.

\subsection{Schematic comparison with \SQB}

A significant portion of the remainder of this paper will involve a
discussion of the possible effects of experimentally-uncertain
nonunique second forbidden transitions (Section~\ref{sec:forbidden})
and the discovery and characterization of convective instability
triggered by thermal conduction
(Section~\ref{sec:onset-electr-capt-24}).  Before discussing these
issues, it is useful to first show a model that encapsulates the
effects of the $A=23$ and $A=25$ isotopes.

\begin{figure}
  \centering
  \includegraphics[width=\columnwidth]{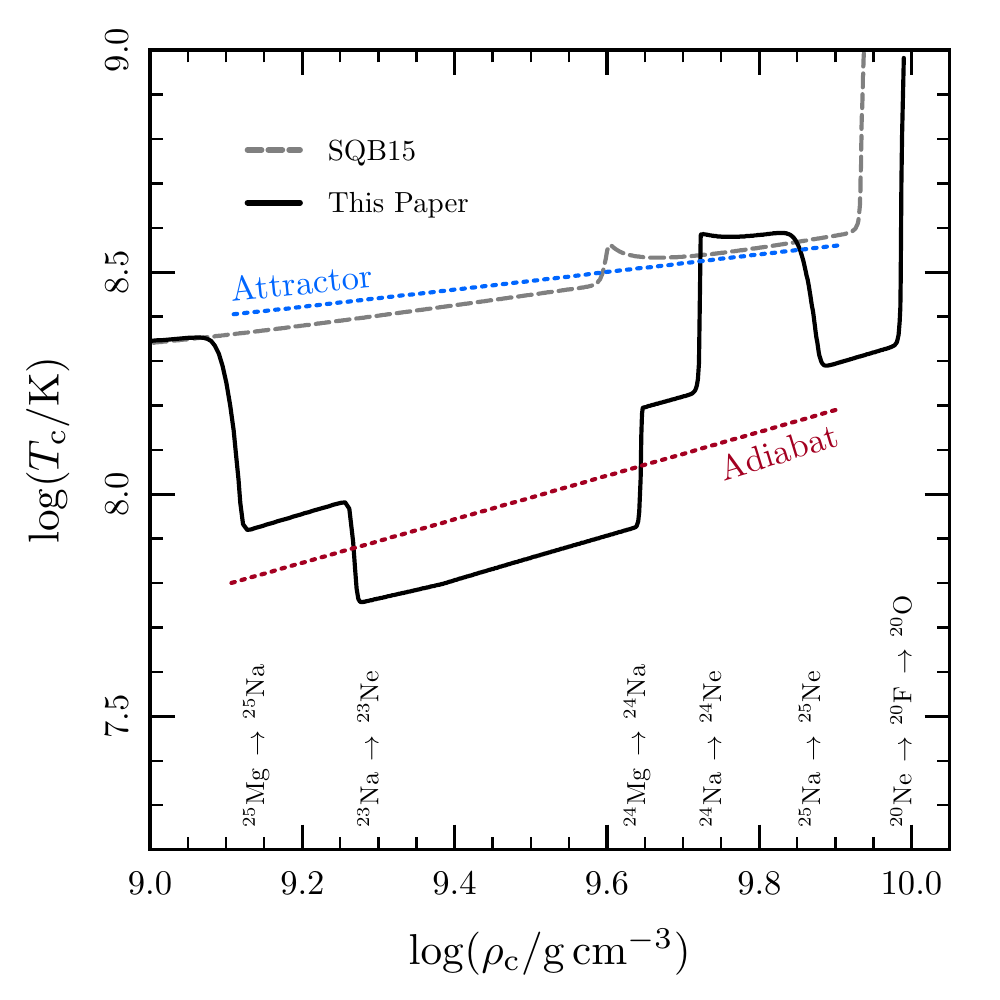}
  \caption{Comparison of a model with (This Paper) and without (SQB15)
    the isotopes \sodium[23] and \magnesium[25].  The key weak
    reactions are indicated at the densities at which they occur,
    accounting only for allowed transitions.  The labeled dotted lines
    show the attractor solution (where neutrino cooling balances
    compressional heating) and a sample adiabat.  These models do not
    include convection which is, however, likely to occur in models with
    significant Urca-process cooling (see Section~\ref{sec:onset-electr-capt-24}).}
  \label{fig:comparison-SQB}
\end{figure}

Fig.~\ref{fig:comparison-SQB} compares the evolution of a model with
the composition used in \SQB\ with a model using a similar composition
but including representative mass fractions of \sodium[23] and
\magnesium[25].  (The precise compositions are given in
Table~\ref{tab:compositions}.)  The models are accreting at a rate of
$\unit[10^{-6}]{\Msunyr}$.  Unless otherwise noted, all models shown
use this fiducial accretion rate.  The model shown in in Fig.~\ref{fig:comparison-SQB} 
neglects forbidden transitions and assumes convective stability and thus
is not the model with the ``best physics''.  However, it ably illustrates
the main point: the evolution of the central temperature is notably
different with Urca-process cooling included.

In \SQB, the temperature immediately prior to electron captures on
\magnesium[24] and \neon[20] was was set by a balance between
compression and neutrino cooling (the attractor solution).  However,
the results in Section~\ref{sec:analytics} demonstrate that for a wide
range of $\Xu$ and $\dot{M}$, significant Urca-process cooling will
occur.  In almost all cases, the WD is cooled to temperatures such
that energy losses by non-nuclear neutrinos \citep[in these conditions
primarily plasma neutrinos, e.g.][]{Itoh1996a} become negligible.
Therefore, we enter regime (i), and expect the material to evolve
along a strongly coupled liquid adiabat.  In these conditions $\Gamma_3 \approx 1.5$
\citep{Chabrier1998}, so $T \propto \rho^{1/2}$.

The difference in threshold density between \sodium[23] (cooling) and
\magnesium[24] (heating) is $\approx 0.4$ dex, and therefore we expect
a temperature increase of 0.2 dex.  This relatively small change in
temperature means that the star does not evolve back onto the
attractor solution before the electron captures on \magnesium[24]
begin.  As shown in Fig.~\ref{fig:comparison-SQB}, the $A=24$ electron
captures begin at a point where the fiducial model has
$\logTc \approx 7.9$.  This affects which electron capture transitions
dominate the rate (see Section~\ref{sec:forbidden24}) and has
implications for the convective stability of this region (see
Section~\ref{sec:onset-electr-capt-24}).  After the energy release
from the $A=24$ electron captures completes, the model evolves back
towards the attractor solution, but around $\logRhoc \approx 9.85$,
additional Urca-process cooling associated with \sodium[25]-\neon[25]
occurs.  In the model shown, this Urca-process cooling is complete well in
advance of the onset of electron captures on \neon[20] (though see
Section~\ref{sec:forbidden20}).

\section{Nonunique second forbidden transitions}
\label{sec:forbidden}

\begin{figure*}
  \centering
  \includegraphics[width=\columnwidth]{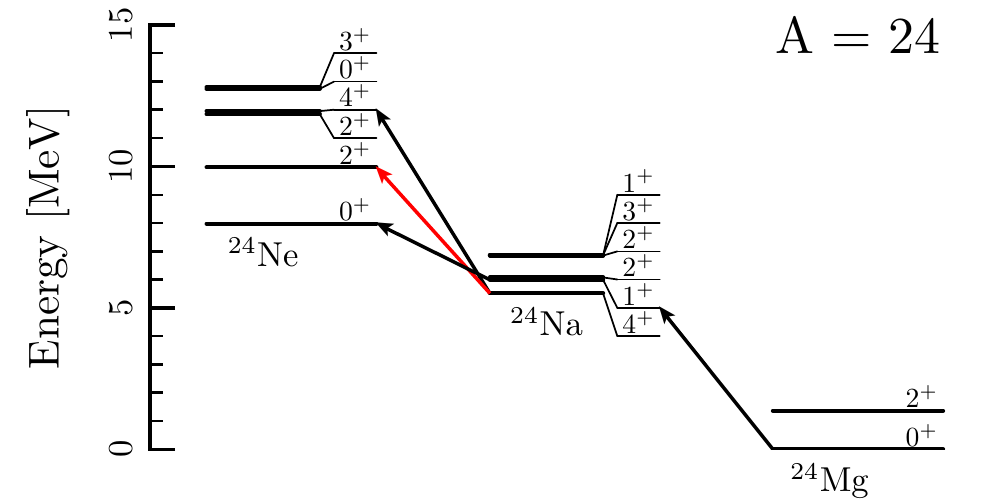}
  \includegraphics[width=\columnwidth]{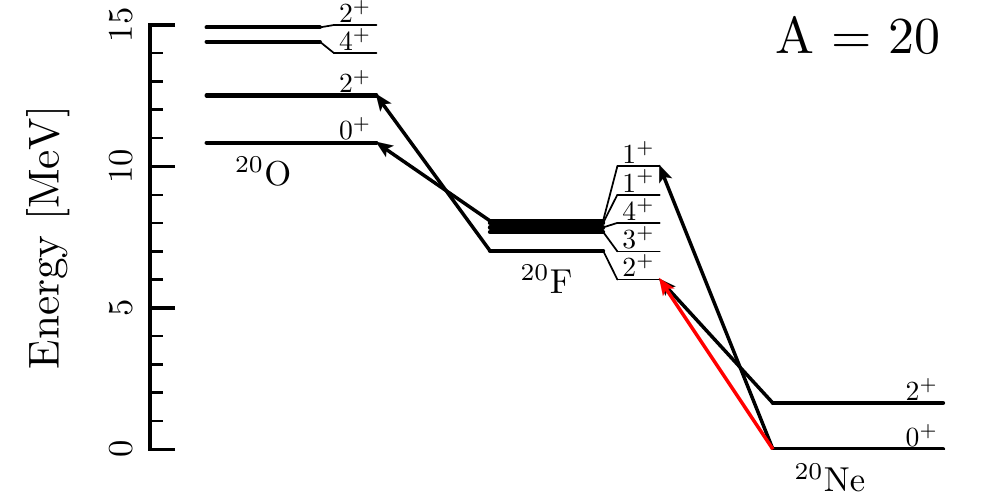}
  \caption{Energy level diagrams for the $A=24$ (left) and $A=20$
    (right) nuclei that we consider.  The $J^{\pi}$ value is indicated
    at the right of each level and is sometimes given an arbitrary
    offset (indicated via a thin line) in order to enhance
    legibility. The most important transitions we consider are
    indicated with arrows.  The red arrows indicate the nonunique
    second forbidden transitions (i.e.  $\Delta J = 2$,
    $\pi_i \pi_f = +1$).}
  \label{fig:levels2420}
\end{figure*}

In Table~\ref{tab:weak-data}, we summarized the key weak reactions and
gave the threshold density associated with the most important
\textit{allowed} transition.  In this section, we discuss the effects
of nonunique second forbidden transitions, which are those with
$\Delta J = 2$, $\pi_i \pi_f = +1$.  Typical $\logft$ values (of
beta decays) for nonunique second forbidden transitions are 11.9-13.6
\citep{Raman1973}.  Such transitions can only dominate the
rate when their threshold density is far enough below the threshold
density of the allowed transition that the additional phase space can allow it to be more rapid.

\citet{MartinezPinedo2014} pointed out the potential importance of the
nonunique second forbidden transition between the ground states of
\neon[20] and \fluorine[20]. The properties of this transition have
not yet been experimentally measured---there exists only an upper
limit \citep{Calaprice1978}---though experiments are being planned
\citep{Kirsebom2017}.  In \SQB, we explored the effect of this
transition and found that while it causes a 0.1 dex shift in the
density at which the initial electron captures on \neon[20] occur, its
effect on the central density at the time of oxygen ignition was more
modest.

There is also a nonunique second forbidden transition between the
ground state of \sodium[24] and the first excited state of \neon[24]
(see left panel of Fig.~\ref{fig:levels2420}).  This transition has a
threshold density below the threshold density for allowed electron
captures from the \sodium[24] ground state.  The effect of this
transition has not previously been explored.\footnote{We thank Gabriel
  Mart\'{i}nez-Pinedo for asking a question about the potential
  importance of such a transition during the Electron-Capture
  Supernovae \& Super-AGB star workshop at Monash in Feb. 2016.}

In this work, we follow \citet{MartinezPinedo2014} in assuming the
phase space shape for these transitions is the same as for the allowed
transitions.  The shape factor for these non-unique transitions can
contain additional powers of the energy, potentially leading to a
factor of 10 increase in the rate for the same $ft$ value
\citep{MartinezPinedo2014}.  Since the $ft$ values for these
transitions is not measured, we present models with different $ft$
values, and this ambiguity is degenerate with our parameter
exploration.

\subsection{Effects for $A=24$}
\label{sec:forbidden24}

The left panel of Fig.~\ref{fig:levels2420} shows the key transitions
for $A=24$.  We now evaluate the relative importance of the
allowed transition from the first excited state of \sodium[24] to the
ground state of \neon[24] and the nonunique second forbidden
transition from the ground state of \sodium[24] to the third excited
state of \neon[24].

We want to evaluate the ratio of these rates at the threshold density
of the $\magnesium[24] \to \sodium[24]$ reaction.  Using an
approximate form of the near-threshold rate \citep[eq. 19
in][]{MartinezPinedo2014} and plugging in the values of the
relevant energy levels and their spins, we find
\begin{equation}
  \frac{\lambda_{\mathrm{forbidden}}} {\lambda_{\mathrm{allowed}}} \approx 0.7  \left[\frac{(ft)_{\mathrm{allowed}}}{(ft)_{\mathrm{forbidden}}}\right]\exp\left(\frac{\unit[0.472]{MeV}}{\kB T}\right)
\end{equation}
For ratios of the $ft$ values in the range $10^{-8}$ -- $10^{-6}$,
this means the forbidden transition dominates when
$T \lesssim \unit[3-4\times10^8]{K}$.  Thus this forbidden transition
was already likely not negligible under the conditions encountered in
\SQB.  In this work, the demonstrated importance of Urca process
cooling means that the temperature at the onset of \magnesium[24]
electron captures is $\logT \la 8$, and thus the forbidden transition is
always important.  Electron captures that proceed via the
forbidden transition deposit more thermal energy per capture into the
plasma; this reflects the difference in the average energy of the
captured electron and emitted neutrino.

However, even though this forbidden transition may dominate the rate,
it might not necessarily be rapid enough that the conversion of
$\sodium[24]$ to $\neon[24]$ will complete before the threshold
density rises to the point that the allowed transition from the
$\sodium[24]$ ground state becomes important.  As indicated in
Table~\ref{tab:weak-data}, this occurs at a threshold density of
$\logRho \approx 9.7$, roughly 0.1 dex above the threshold density for
$\magnesium[24]$.

Using the compression time estimate from
equation~\eqref{eq:trho-numerical}, it will take approximately
$\unit[5\Mdot_{-6}^{-1}]{kyr}$ to achieve this density change.
Therefore, electron captures via the allowed transition will be
important when the reaction time-scale
$\lambda_{\mathrm{forbidden}}^{-1}$ is longer than this compression
time-scale.  This corresponds to the approximate condition
$\logft \gtrsim 15 - \log(\Mdot_{-6})$.  Thus, for the expected
$ft$ value, the electron captures on $\sodium[24]$ will typically
not be delayed to higher density.

\subsection{Effects for $A=20$}
\label{sec:forbidden20}

The right panel of Fig.~\ref{fig:levels2420} shows the key
transitions for $A=20$.  \citet{MartinezPinedo2014} report that for
densities in the range $9.6 \la \logRho \la 9.9$, the forbidden
transition dominates the rate for $\logT \la 8.8$ (assuming the
forbidden transition strength is at its experimental upper limit).

Thus again, while the forbidden transition may dominate the rate,
there is not necessarily time for substantial $\neon[20]$ captures
before the threshold density associated with the allowed transition
occurs. Performing a similar estimate as in the $A=24$ case gives the
approximate condition that the electron captures on the allowed
transition will be important when
$\logft \gtrsim 13 - \log(\Mdot_{-6})$.  This estimate is consistent
with the results reported in \SQB.  Given that threshold density for
the \sodium[25]-\neon[25] Urca pair is $\logRho \approx 9.8$, when
this forbidden transition is important we expect both exothermic
$A=20$ electron captures and $A=25$ Urca-process cooling to be
operating at the same location in the star.

\subsection{Exploration of effects}
\label{sec:forbiddenexploration}

\begin{figure}
  \centering
  \includegraphics[width=\columnwidth]{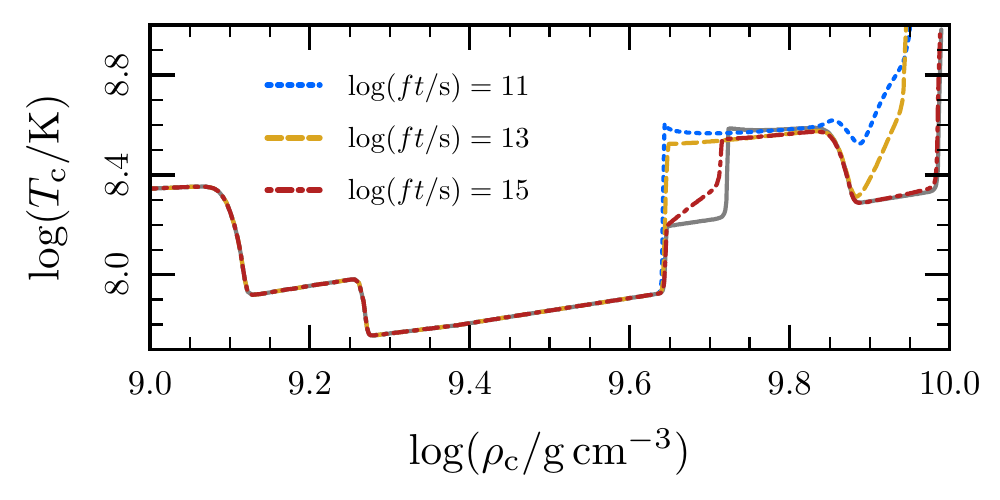}
  \caption{Models with a range of $ft$ values for the uncertain nonunique
    second forbidden transitions.  The grey line shows a
    model which only includes allowed transitions.  At this accretion
    rate, $\Mdot = \unit[10^{-6}]{\Msunyr}$, the nonunique second
    forbidden transitions are only negligible if $\logft \ga 15$.}
  \label{fig:comparison-logft}
\end{figure}

To explore their effects, we vary the strength of the nonunique second
forbidden transitions.  For convenience, we choose the two transitions
to have the same \textit{beta-decay} $ft$ value.  We run models with
values $\logft = $ 11, 13 and 15, setting the $ft$ values for electron
capture correspondingly, including the ratio of the spin degeneracies.
There is no physical reason that the
transitions need to have the same strength.  However, for the models
shown in Fig.~\ref{fig:comparison-logft}, the star returns to the
``attractor'' solution between the $A=24$ and $A=20$ electron
captures, largely erasing the previous effects.  Therefore, one can
roughly assess the effects of each transition independently.

The tracks shown in Fig.~\ref{fig:comparison-logft} agree with the
estimates in the previous subsections as to when the each of the
transitions is important.  The nonunique second forbidden transition
in \sodium[24]-\neon[24] determines whether the \sodium[24]
electron captures occur immediately after those on \magnesium[24] or
whether they are delayed to higher density, but does not appear to
have an effect on the subsequent evolution.  However, given the
important role that the $A=24$ electron captures play in the onset of
convective instability (see Section~\ref{sec:onset-electr-capt-24}),
such a delay could in principle have an effect that would not be
revealed by the models in this paper.  When the nonunique second
forbidden transition in \neon[20]-\fluorine[20] is important
($\logft \la 13$), we do not see a significant dependence of the
ignition density on the strength of the transition.  When this
transition is unimportant ($\logft \approx 15$), then the Urca-process
cooling by \sodium[25]-\neon[25] at $\logRho \approx 9.85$
significantly cools the material and leads to electron captures on
$\neon$ that begin at slightly higher density ($\approx 0.05$ dex)
than the other models.

\section{Stability during and after the electron captures on \magnesium[24] and \sodium[24]}
\label{sec:onset-electr-capt-24}

Previous models of accreting ONe WDs have found that the WD remains
convectively stable when using the Ledoux criterion for convection
\citep{Miyaji1980, Miyaji1987, Canal1992, Hashimoto1993, Gutierrez1996,
  Gutierrez2005}.  Even though the entropy release from the electron
captures creates a highly superadiabatic temperature gradient, it does
not trigger convection because of the stabilizing $\Ye$-gradient.

\begin{figure}
  \centering
  \includegraphics[width=\columnwidth]{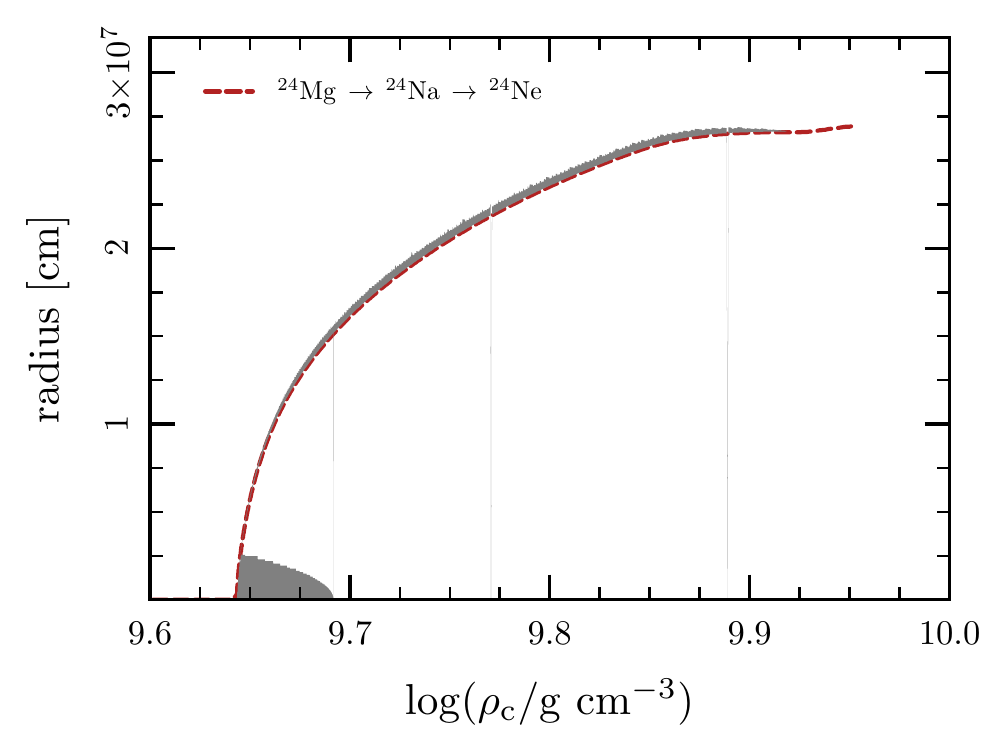}
  \caption{Location of convectively unstable regions in the fiducial
    model.  On the $x$-axis, $\logRhoc$ serves as a proxy for time.
    The dashed line shows the location of the $A=24$ electron capture
    front (defined by the place where the \magnesium[24] and \neon[24]
    abundances are approximately equal).  The grey shaded regions show
    where $N^2 < 0$. The action of convection has been artificially
    suppressed in this model; no mixing occurs in convectively
    unstable regions.}
  \label{fig:unstable_regions}
\end{figure}

The \MESA\ models in this paper, which are the first to include the
effects of significant Urca-process cooling, do develop regions of
convective instability.  In regions where the electron captures are
occurring, and thus where the temperature and \Ye\ gradients are
necessarily tightly linked, the material remains convectively stable,
consistent with previous results.  However, our models show the
development of convectively unstable regions (i.e. where $N^2< 0$) in
the core after the $A=24$ electron captures have completed and also
off-centre, ahead of the region where the $A=24$ electron captures are
occurring.  Fig.~\ref{fig:unstable_regions} shows the convectively
unstable regions in our fiducial model.  In these models, we evaluate
convective instability via the Ledoux criterion, but suppress the
action of convection once unstable regions develop (see
Section~\ref{sec:mesa-options} for the \MESA\ options used).

These results indicate that the temperature of the material
immediately before it undergoes the $A=24$ electron captures has a
profound effect on the convective stability of the model.  This
temperature dependence is a consequence of the steep temperature
dependence of the electron capture reactions and the subsequent
influence of thermal conduction.

As material in the centre of the WD nears the threshold density, the
electron-capture reaction rates increase and the reactions will
proceed in earnest once the reaction time-scale is of order the
compression time-scale.  This happens while the reaction is still
sub-threshold, meaning the reaction rate has an exponential dependence
on the temperature.  This strong temperature dependence allows for a
thermal runaway.

The initial length scale for the runaway is set by hydrostatic
equilibrium.  The core is approximately isothermal, but the pressure
decreases with increasing radius.  This implies a gradient in the
electron chemical potential and thus a length scale over which the
electron capture rate (and hence the heating rate) varies by order
unity.  Recall that in the sub-threshold limit the rate varies with
temperature as
\begin{equation}
  \lambda \propto \left(\frac{\kB T}{\me c^2}\right)^3 \exp\left(\frac{\mu + Q}{\kB T}\right) ~.
\end{equation}
So the rate varies by a factor of $e$ over a length scale where the
chemical potential changes by $\Delta \mu = k_B T$.  Since the thermal
contribution to the pressure is not significant, the variation of the
chemical potential is roughly independent of the temperature, meaning the
length scale of this variation in the rate is smaller at lower
temperatures.

Additionally, at a lower initial temperature, for the reaction rate to
reach a given value, the material must get closer to the threshold
density (i.e. $(\mu + Q)/(\kB T)$ will be less negative).  This means
that once the temperature rises as a result of the heating, the
reaction rate is faster at a given temperature.  This means that the
thermal runaway will complete in a shorter amount of time.

\begin{figure}
  \centering
  \includegraphics[width=\columnwidth]{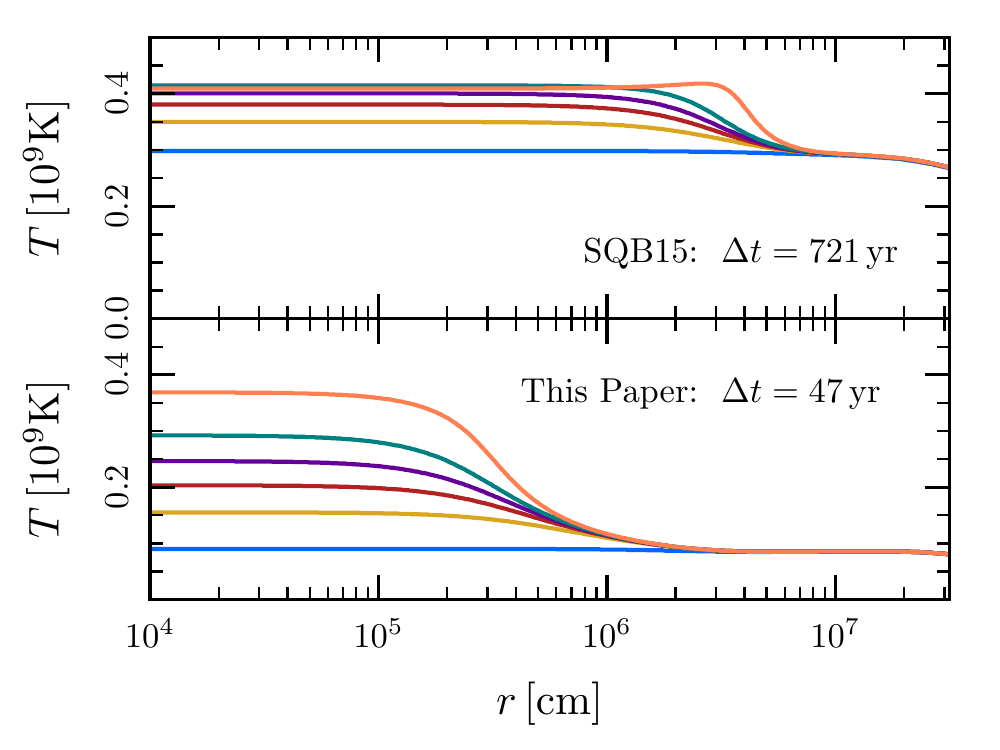}
  \caption{Temperature profiles during the thermal runaway caused by
    the $A=24$ electron captures.  The top panel shows a model that
    did not include earlier Urca-process cooling and hence is
    significantly hotter than the bottom panel which did include
    Urca-process cooling.  Profiles are shown at times when the
    central $\magnesium[24]$ mass fraction is 0.0495, 0.04, 0.03,
    0.02, 0.01, and 0.0005. (The temperature increases as
    $\magnesium[24]$ is consumed.)  The annotation indicates the time
    elapsed between first and last profile shown.  The runaway that
    begins at an initially colder temperature has a smaller length
    scale and time-scale. None of the profiles shown are convectively
    unstable, though instability will shortly set in for the model in
    the bottom panel.}
  \label{fig:runaway}
\end{figure}

Fig.~\ref{fig:runaway} shows temperature profiles during the runaway
for a model that has not experienced Urca-process cooling (top) and
one that has (bottom).  The two most important differences are readily
apparent: the length scale of the initially colder runaway is
significantly smaller and the runaway completes in substantially less
time.  In the \SQB\ model, the longer time means that thermal neutrino
cooling is  more important; this effect explains why the central
temperature is not at the maximum temperature in the final profile shown in the top panel.

The thermal runaway associated with the $A=24$ electron captures ends
because of the exhaustion of \magnesium[24].  This is the key
difference between this runaway and the \neon[20] runaway that we
studied in detail in \SQB, where thermonuclear oxygen begins before
\neon[20] depletion.  As the runaway ends and each parcel reaches the
post-capture composition, no residual composition gradient remains.
However, a residual temperature gradient does remain because of the
slight gradient in the reaction rate and the differential effects of
thermal conduction.  This residual temperature gradient is
superadabiatic and thus the material is convectively unstable.  This
produces a central convection zone on the length scale of the initial
thermal runaway, as shown in Fig.~\ref{fig:unstable_regions}.  In
Appendix~\ref{sec:toy-runaway}, we reproduce this behavior in a simple
toy model.

The thermal runaway produces a small hotspot at the centre of the
star.  Heat from this hotspot will be conducted outwards, but this
cannot lead to the propagation of the electron-capture front through a
significant portion of the star, since the reactions only occur in
material near or above the threshold density.  Therefore, as in
previous models, the electron-capture front moves outwards as a consequence of
accretion.

The front is advancing slowly, moving through the star on the
compression time-scale, and so conduction can move heat ahead of it.
Ahead of the front, where the chemical potential is lower, the
electron-capture reactions are sufficiently slow that an increase in
the temperature does not result in significant composition change.
Thus in these regions a temperature gradient develops without a
corresponding $\Ye$ gradient and the region becomes convectively
unstable.  Models that have experienced Urca process cooling are more
prone to experience this because their steeper temperature
gradients favor conduction and because with a lower upstream
temperature less heat is required to get a super-adiabatic temperature
gradient.

\begin{figure}
  \centering
  \includegraphics[width=\columnwidth]{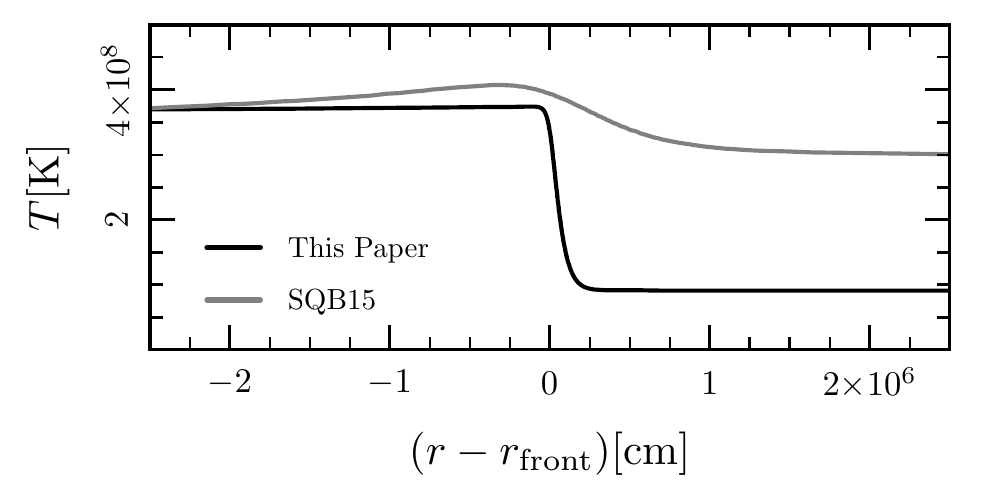}
  \caption{Temperature profiles near the $A=24$ electron-capture front
    in the fiducial model and the model from \SQB\ (shown at the time when
    $\logRhoc = 9.7$).  The lower upstream temperature due to Urca
    cooling causes the temperature gradient to be much steeper,
    enhancing the destabilizing effects of thermal conduction.}
  \label{fig:capture_front}
\end{figure}

If we allow convection to operate via the usual mixing length theory
(MLT), it becomes extremely difficult for \MESA\ to proceed and we
have not been able to construct numerically-converged \MESA\ models
beyond this point.  When a zone becomes convectively unstable, its
temperature gradient changes from the radiative gradient to the
adiabatic gradient.  These temperature changes affect the structure in
a way that alters the convective stability of other zones.  As \MESA\
iterates to find the new solution, the convective boundaries change at
each iteration and the solver fails to converge.\footnote{We have
  encountered related difficulties in \citet{Brooks2017} and continue
  to work to understand how to improve the situation in \MESA.}

\begin{figure}
  \centering
  \includegraphics[width=\columnwidth]{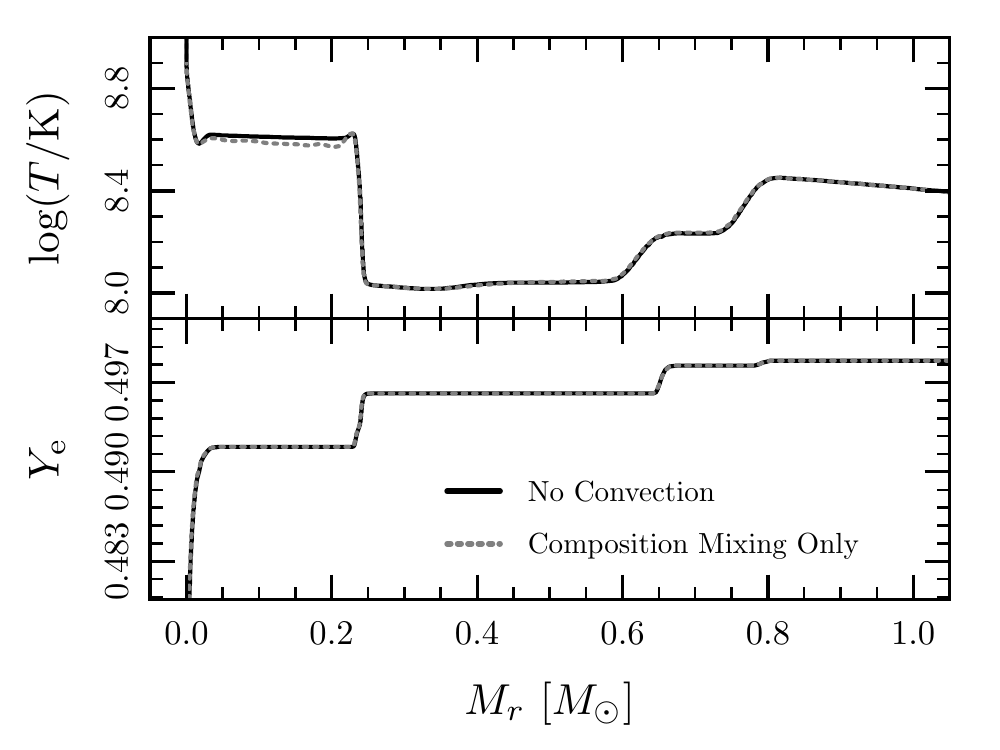}
  \caption{Comparison of temperature (top panel) and composition
    (bottom panel) profiles for models at the time of oxygen ignition.
    The model with no convection and the model in which convection
    only mixes composition (but does not modify temperature) agree
    well.  This indicates that if the convection zones do not grow,
    they will not substantially affect the evolution of the models.}
  \label{fig:mixing}
\end{figure}

As a demonstration that it is the temperature change that causes the
problem (at least initially), and not the composition mixing, we ran a
model that does
not allow MLT to change the temperature, but retains the normal MLT
diffusion coefficients for mixing of composition (see Section~\ref{sec:mesa-options} for the \MESA\ options used).  Fig.~\ref{fig:mixing} compares this model
(at the time of oxygen ignition) with our fiducial model in which
convection is completely suppressed.  No significant difference exists
between the two models.  Physically, the regions that become unstable
are ahead and behind the electron-capture front in regions that do not have
substantial composition gradients.  Thus composition mixing in these
regions cannot by itself have a significant effect.

The essential question that must be answered about these convectively
unstable regions is: do they want to grow?  In particular, do they
grow and ultimately lead to the formation of a long-lived central
convection zone?  Previous work has demonstrated the qualitative
difference between models that develop a convective core and those
that don't \citep{Miyaji1980, Miyaji1987}.\footnote{In previous work,
  models that developed convective cores were those in which stability
  was evaluated using the Schwarzchild criteron, which we do not think is appropriate.}  The presence of a
central convection zone means the heating from the electron captures
is effectively deposited over the entire convective region.  With a
greater mass to heat, more material must undergo electron captures to
cause the core to reach conditions for oxygen ignition.  Models with
central convection zones at the onset of \neon[20] captures do not
reach oxygen ignition until much higher densities, strongly favoring
their collapse to form a neutron star.

However, the evolution of models with large central convection zones
is subject to the substantial uncertainties associated with the
\textit{convective} Urca process \citep{Paczynski1972b}.  Once the convection zone grows to span the
threshold density of one or more of the Urca pairs, convective motions
can transport material that has undergone electron captures in higher
density regions to lower density regions where it will beta decay (and
vice-versa).  The greater abundances of Urca-pair isotopes in ONe WDs
(compared to CO WDs) will increase the importance of this process.
The interaction of the convective mixing and the reactions is
difficult to model.  The development of a treatment for the convective
Urca process and its effects suitable for inclusion in stellar
evolution codes remains an active area of research \citep[e.g.][]{
  Lesaffre2005a}.

Therefore, it is non-trivial but of critical importance to explore the outcome of
these convectively unstable regions and how to best model
them in stellar evolution codes.  We necessarily defer this difficult
problem to future work.

\section{Subsequent evolution towards collapse}
\label{sec:subsequent-evolution}

Section~\ref{sec:onset-electr-capt-24} demonstrated the onset of
localized convective instability after the $A=24$ electron captures
begin.  Uncertainties in how to treat the evolution at this point mean
that the later evolution is necessarily less certain.  For now, we
proceed by artificially suppressing the action of convection.  This
allows us to characterize the evolution of
models in which a long-lived central convection zone does not develop.

\subsection{Onset of electron captures on \neon[20] and \fluorine[20]}
\label{sec:onset-electr-capt-20}

As discussed in \SQB, the electron captures on $\neon$ trigger a
thermal runaway that leads to the formation of an outgoing oxygen
deflagration wave.  The final fate of the star is determined by a
competition between the propagation of the oxygen deflagration and
electron captures on the material (in nuclear statistical equilibrium;
NSE) behind the deflagration front \citep{Nomoto1991}.  The speed of
the deflagration and the electron capture rate on its NSE ash are both
functions of density and electron fraction.  \citet{Timmes1992} found
that the deflagration speed scaled $\propto \rho^{1.06}$. At the
relevant densities, the neutronization time-scale scales roughly as
$\rho^{-0.5}$ \citep[see figure 13 in \SQB\ and][]{Seitenzahl2009}.
Studies of the final fate of these objects typically explore
uncertainties in the initial models by varying the central density at
oxygen ignition \citep[e.g.][]{Jones2016c}.  Therefore, we now
describe the range of central densities found in our models.

The temperature affects the density at which the electron captures on
\neon[20] begin, with lower temperatures corresponding to higher
densities (see figure 4 in \SQB), so Urca-process cooling can
influence the ignition density.  In Fig.~\ref{fig:comparison-logft},
we showed that if the nonunique second forbidden transition is
unimportant ($\logft \approx 15$), then the Urca-process cooling by
\sodium[25]-\neon[25] at $\logRho \approx 9.85$ effectively sets the
temperature at which the electron captures on $\neon[20]$ begin.  This
leads to electron captures on $\neon$ that begin at slightly higher
density ($\approx 0.05$ dex) than in \SQB.  In cases where this
forbidden transition is important ($\logft \la 13$), we do not see a
significant dependence of the ignition density on the details of the
transition.

\begin{figure}
  \centering
  \includegraphics[width=\columnwidth]{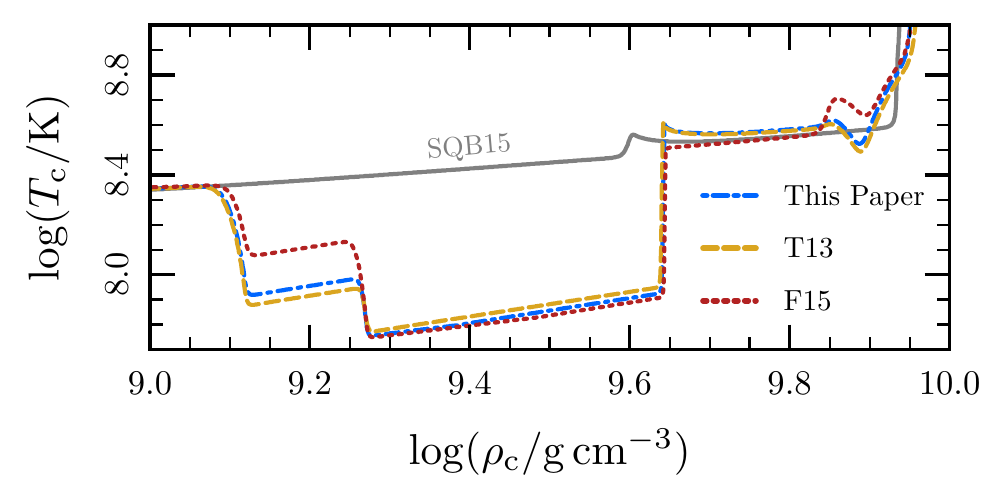}
  \caption{Comparison of a model with our fiducial composition (This
    Paper) with two compositions based on recent results of the
    evolution of intermediate mass stars:
    \citep[T13;][]{Takahashi2013} and \citep[F15;][]{Farmer2015}. The
    solid line shows the model from \SQB.  The precise compositions
    are given in Table~\ref{tab:compositions}.  While the details of
    the evolution depend on the abundances, the density at which
    electron captures on $\neon[20]$ trigger the oxygen deflagration
    appears to be insensitive to the precise composition.  The density
    is only slightly higher $(\approx 0.02$ dex) than the model in
    \SQB.  All models neglect convection.}
  \label{fig:comparison-all}
\end{figure}

The composition could also influence the ignition density.  We do not
perform an extensive parameter study, but
Fig.~\ref{fig:comparison-all} shows the evolution of the central
density and temperature for the representative compositions listed in
Table~\ref{tab:compositions}.  We see minor differences between the
evolutionary tracks.  For example, the F15 models have the lowest
abundance of $A=25$ elements and this accounts for the differences in
cooling around $\logRho \approx 9.1$ and $9.85$.  However, the density
at which electron captures on $\neon[20]$ trigger oxygen ignition
is insensitive to the precise details of the composition.

\subsection{Propagation of the oxygen deflagration}
\label{sec:oxygen-deflagration}

Independent of their cooling effects, the electron captures on
\sodium[23], \magnesium[25], and \sodium[25] have reduced the $\Ye$ of
the material.  For the fiducial composition, this change is
$\Delta \Ye \approx -3 \times 10^{-3}$. A reduction in $\Ye$ increases
both of the oxygen deflagration speed and the electron-capture rates on the oxygen burning ashes.  \citet{Timmes1992} found that
reducing $\Ye$ from 0.50 to 0.48 reduced the deflagration speed by
approximately 30 per cent.  In the tabulated electron-capture rates on
NSE material from \citet{Seitenzahl2009}, changing $\Ye$ from 0.50 to
0.48 at $\logRho \approx 9.9$ and $\logT \approx 10$ increases the
neutronization time-scale by approximately a factor of 2.5. Note that
these changes are quoted for a $\Delta \Ye$ approximately 10 times
greater than the difference here.

The exact competition between these two processes is best probed via
simulations which can include both the physics of the oxygen
deflagration and the NSE electron captures.  However, the changes due
to a possible increase in density and the decrease in $\Ye$ are
relatively small and in opposite directions; they are unlikely to
significantly affect the fate of the outwardly-going oxygen flame.

\subsection{Other effects of reduced electron fraction}
\label{sec:other-ye-effects}

The electron captures on the A=23 and A=25 isotopes reduce $\Ye$ in
the material in the WD that has exceeded the threshold density for
these reactions.  At oxygen ignition, this has occurred in about half
of the star (see Fig.~\ref{fig:mixing}).  The Chandrasekhar mass
scales with $\Ye^2$, and so at the onset of collapse, models which
experience this reduction in \Ye\ will have lower masses relative to
models in which these composition shifts have not been accounted for.

The models shown in Fig.~\ref{fig:comparison-SQB} have different
masses at the time of the formation of the oxygen deflagration (and
hence the likely collapse to a NS).  The mass difference between these
two models is $\approx 0.016$ \Msun, with the model that included the
odd mass number isotopes having the lower mass.  Studies that use the
observed mass of low-mass neutron stars (thought to be formed via AIC
or electron-capture supernova) to make inferences about the nuclear
equation of state \citep[e.g.][]{Podsiadlowski2005} require knowing
the baryonic mass of the WD just prior to collapse.  A mass difference
of $\unit[0.01]{\Msun}$ is the same order of magnitude as the effects
of finite temperature, general relativity, and Coulomb corrections,
which are important in formulating such constraints. To realize the
suggestion of \citet{Podsiadlowski2005} that one can ultimately
pinpoint the baryonic mass of the core to within
$\unit[2\times10^{-3}]{\Msun}$ will require realistic temperature and
composition profiles.

\section{Effect of accretion rate}
\label{sec:mdot}

In Sections~\ref{sec:mesa-calculations}-\ref{sec:subsequent-evolution}
we focused on models accreting at a constant rate of
$\Mdot = \unit[10^{-6}]{\Msunyr}$. Near the Chandrasekhar mass
$(\ga 1.3 \Msun)$, the range of mass accretion rates for
thermally-stable hydrogen burning is
$\approx \unit[4-7\times10^{-7}]{\Msunyr}$ \citep{Wolf2013} and for
thermally-stable helium burning is
$\approx \unit[1.5-4.5\times10^{-6}]{\Msunyr}$
\citep{Brooks2016}. Thus our fiducial choice represents an accretion
rate that is approximately characteristic of any
stably-burning accretor.  However, it is useful to repeat the models
for a range of accretion rates; such a parameter study was presented
in \SQB\ and we now update that result including the effects of the Urca
process.

\begin{figure}
  \centering
  \includegraphics[width=\columnwidth]{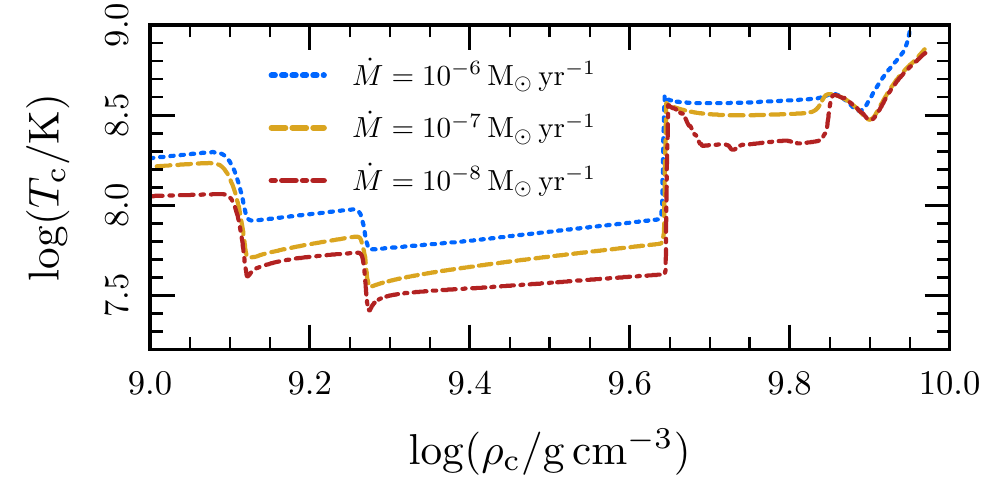}
  \caption{Central density-temperature trajectories of models with different accretion rates.
    Models with lower accretion rates have lower temperature, but the
    overall evolution is similar.}
  \label{fig:mdots}
\end{figure}

As discussed in Section~\ref{sec:analytics}, at lower accretion rates,
the WD will be cooler.  This is because the longer compression
time-scale means the balance between compressional heating and thermal
neutrino losses occurs at lower temperature; additionally, once the
Urca-process neutrino cooling occurs it will cool material to a lower
temperature.  Fig.~\ref{fig:mdots} shows the evolution of the central
conditions for models with several accretion rates and both of these
effects are evident.

\begin{figure}
  \centering
  \includegraphics[width=\columnwidth]{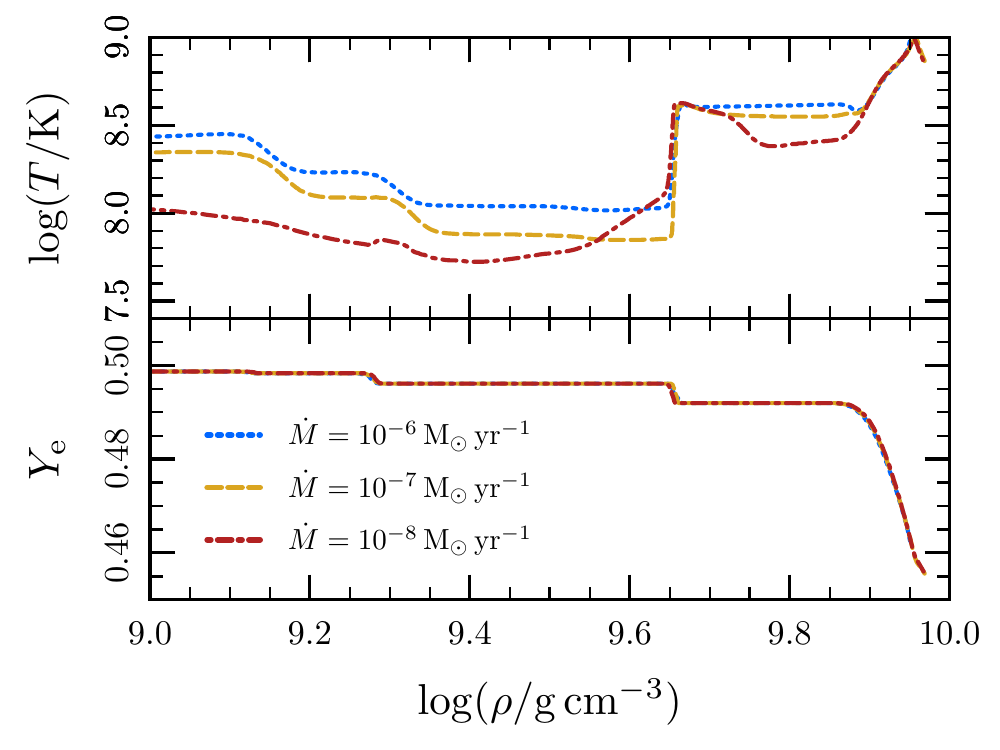}
  \caption{Temperature (top panel) and composition
    (bottom panel) profiles at the time of oxygen ignition for models
    with different accretion rates.}
  \label{fig:mdots-profiles}
\end{figure}

In a cooler WD, the physical width of the regions over which the weak
reactions primarily occur will be narrower (since the extent scales
$\propto \kB T / \EF$).  Both the longer compression time-scale and the
shorter lengthscale serve to enhance the relative importance of
thermal conduction. Fig.~\ref{fig:mdots-profiles} plots $T$ and $\Ye$
profiles for the models shown in Fig.~\ref{fig:mdots} at the time of
oxygen ignition.  The effects of thermal diffusion can be seen in the
shallower temperature gradients.  This is particularly easy to see
around $\logRho \approx 9.6$ in the model with
$\Mdot = \unit[10^{-8}]{\Msunyr}$, where it is evident that
substantial heat from the $A=24$ electron captures has diffused to
lower density.  Consistent with this fact, in models with lower
accretion rates we observe larger regions that are convectively
unstable due to the effect discussed in
Section~\ref{sec:onset-electr-capt-24}.

As noted in \SQB, the non-unique second forbidden transition can lead
to mildly off-centre ignitions if its strength is near the
experimental upper limit (see figure 12 and surrounding discussion).
As can be seen in Fig.~\ref{fig:mdots} and
Fig.~\ref{fig:mdots-profiles}, the models with $\Mdot$ of
$\unit[10^{-7}]{\Msunyr}$ and $\unit[10^{-8}]{\Msunyr}$
experience mildly off-centre ignitions with the fiducial transition
strength of $\logft =11$.  This shows that at a fixed transition
strength, lower accretion rates lead to off-centre ignitions. When an
off-centre ignition does occur, the ignition location is
$\la \unit[50]{km}$ from the centre of the WD.  Given the
uncertainties in the strength of the non-unique second forbidden
transition, we defer a more thorough characterization of this effect
to future work.

\section{Conclusions}
\label{sec:conclusions}

We have demonstrated the substantial effects that Urca-process cooling
has on the thermal evolution of accreting ONe WDs.  We have provided a
simple analytic expression for the peak Urca-process cooling rate
(equation~\ref{eq:urca-C-max}) and used it to derive an approximate
expression for the temperature to which the Urca process cools the
plasma (equation~\ref{eq:t-urca}).  We used a suite of \MESA\
simulations to confirm these simple analytic scalings
(Figs.~\ref{fig:scalings-logx} and \ref{fig:scalings-mdot}).  The
magnitude of these effects is inconsistent with earlier work by
\citet{Gutierrez2005}, who severely underestimate the amount of
Urca-process cooling.

As discussed by \citet{Paczynski1973}, Urca-process cooling will also
occur in accreting CO WDs, where it leads to an increase in the
density at which carbon is ignited.  This effect has not been fully
explored in the context of Type Ia supernova progenitors.  The
estimates we provide in Section~\ref{sec:analytics} are equally
applicable in this case \citep{Denissenkov2015, MartinezRodriguez2016,
  Piersanti2017}

In Section~\ref{sec:forbidden} we characterized the effects of two
nonunique second forbidden transitions.  Since the strength of these
transitions has not yet been experimentally measured, we characterized
their effect for a range of transition strengths
(Fig.~\ref{fig:comparison-logft}).  One transition, in
\neon[20]-\fluorine[20], has been previously discussed by
\citet{MartinezPinedo2014} and \SQB.  In this paper we showed that
this transition is important at the same density where cooling from
the \sodium[25]-\neon[25] Urca pair is occurring.  The other
transition, in \sodium[24]-\neon[24], has not previously been
discussed; we find it does appear to be important in determining the
rate.  Given the role the $A=24$ electron captures play in causing
convective instability, it would be desirable to better measure this
transition strength.

In Section~\ref{sec:onset-electr-capt-24} we showed that Urca-process
cooling has another important consequence.  It leads to lower
temperatures at the onset of $A=24$ captures in turn producing
convectively unstable regions, even when using the Ledoux criterion.
In Section~\ref{sec:onset-electr-capt-24} and
Appendix~\ref{sec:toy-runaway}, we explained how thermal conduction
leads to this outcome.  Numerical difficulties associated with the
development of these convectively unstable regions prevented us from
evolving the models further while modeling convection using normal
mixing length theory.  We showed that if the convection zones mix only
localized regions, their effect on the subsequent evolution is
minimal.  However, if these convection zones were to grow to encompass
a significant fraction of the star, their effect on its evolution
would be profound; models that have large convective cores undergo
collapse at significantly higher density \citep{Miyaji1980}.
Understanding the dynamics of these convection zones will be an
important avenue for future work.  Multi-dimensional hydrodynamics
simulations may be able to help determine whether these convection
zones want to grow.  Useful results may also be obtained from stellar evolution calculations using mixing prescriptions that circumvent 
the numerical difficulties encountered in this work.

In Section~\ref{sec:subsequent-evolution} we continued to evolve our
models up to the onset of oxygen ignition, under the assumption that
the convectively unstable regions do not substantially alter the
structure of the WD.  We find similar central densities at the time of
oxygen ignition as \SQB.  This suggests that inclusion of Urca-process
cooling does not affect the conclusion that the final outcome of
accreting ONe WDs approaching the Chandrasekhar mass is
accretion-induced collapse to a neutron star \citep{Nomoto1991}.
However, this conclusion is provisional given the uncertainties
introduced by convection.  In addition, recent multi-dimensional work
has begun to revisit the critical density threshold
\citep{Jones2016c}.  Future work using hydrodynamical models and
realistic progenitor models can help elucidate whether aspects such as
the different $\Ye$ profiles have an effect on the collapse.

\section*{Acknowledgements}

We thank Evan Bauer, Jared Brooks, Rob Farmer, Daniel Lecoanet, Ken'ichi Nomoto, Bill
Paxton, Philipp Podsiadlowski, Frank Timmes, and Bill Wolf for useful
discussions.  We thank Toshio Suzuki for providing machine-readable versions of the tables from \citet{Suzuki2016}.
We thank the anonymous referee for a helpful report.
We acknowledge stimulating workshops at Sky House where
these ideas germinated.  Support for this work was provided by NASA
through Hubble Fellowship grant \# HST-HF2-51382.001-A awarded by the
Space Telescope Science Institute, which is operated by the
Association of Universities for Research in Astronomy, Inc., for NASA,
under contract NAS5-26555.  JS was also supported by the NSF Graduate
Research Fellowship Program under grant DGE-1106400 and by NSF grant
AST-1205732.  LB is supported by the National Science Foundation under
grant PHY 11-25915.  This research is funded in part by the Gordon and
Betty Moore Foundation through Grant GBMF5076 to LB and EQ.  EQ is
supported in part by a Simons Investigator award from the Simons
Foundation and the David and Lucile Packard Foundation.  This research
used the SAVIO computational cluster resource provided by the Berkeley
Research Computing program at the University of California Berkeley
(supported by the UC Chancellor, the UC Berkeley Vice Chancellor of
Research, and the Office of the CIO).  This research has made use of
NASA's Astrophysics Data System.


\bibliographystyle{mnras}
\bibliography{Urca}


\appendix

\section{Maximum Urca Cooling Rate}
\label{sec:urca-physics}

The expressions for the rates of electron-capture and beta-decay
reactions have been previously derived
\citep[e.g.][]{Tsuruta1970,Fuller1985,MartinezPinedo2014}.  In this
Appendix, for completeness, we give expressions for these rates,
specialized to the Urca process, with the goal of extracting a simple
expression for the maximum Urca-process cooling rate.  We consider
only the allowed ground state to ground state transition of an Urca
pair.  We choose the isotope undergoing electron capture to have
charge $Z$ and thus the isotope undergoing beta decay has charge
$Z-1$.  We always assume the electrons are relativistic with energy
$\Ee \gg \me c^2$.

The rate of electron capture or beta decay can be written as
\begin{equation}
  \label{eq:lambda}
  \lambda = \frac{\ln 2}{(ft)} I(\mu, T, Q),
\end{equation}
where $ft$ is the comparative half-life (typically given in units of
seconds) and can be either measured experimentally or theoretically
calculated from the weak-interaction nuclear matrix elements.  The phase space factor
$I$ depends on the temperature $T$, electron
chemical potential $\mu$, and the energy difference between the parent
and daughter states $Q$.  The value of $Q$ includes both the nuclear
rest mass and the energy associated with excited states.  Similarly,
the rate of energy loss via neutrinos is
\begin{equation}
  \label{eq:epsilon}
  \varepsilon_\nu = \frac{ \me c^2 \ln 2}{(ft)} J(\mu, T, Q)~,
\end{equation}
where $J$ is a phase space factor that contains an additional power of
the neutrino energy.

For convenience, we define $\beta = (k_B T)^{-1}$ and the
non-dimensionalized parameters $q = \beta |Q|$,
$\theta = \beta \me c^2$, $\eta = \beta \mu$, $\epsilon = \beta \Ee$.
The value of $I$ for electron capture is
\begin{equation}
I_\mathrm{ec} = \theta^{-5} \exp(\upi\alpha Z)\int_{q}^{\infty} \frac{\epsilon^2 (\epsilon - q)^2}{1 + \exp(\epsilon - \eta)} d\epsilon~,
\end{equation}
and the value of $J$ for electron capture is
\begin{equation}
J_\mathrm{ec} = \theta^{-6} \exp(\upi\alpha Z) \int_{q}^{\infty} \frac{\epsilon^2 (\epsilon - q)^3}{1 + \exp(\epsilon - \eta)} d\epsilon~,
\end{equation}
where $\alpha$ is the fine structure constant.  These integrals can
easily be rewritten (using the substitution $x = \epsilon - q$) in
terms of the complete Fermi integrals, which are defined as
\begin{equation}
  \label{eq:fdintegral}
  F_k(y) = \int_0^{\infty} \frac{x^k}{1 + \exp(x-y)} dx~.
\end{equation}
Doing so gives
\begin{equation}
I_\mathrm{ec} = \theta^{-5} \exp(\upi\alpha Z) \left[ F_4(\delta) + 2 q F_3(\delta) + q^2 F_2(\delta) \right]~,
\end{equation}
and
\begin{equation}
J_\mathrm{ec} = \theta^{-6} \exp(\upi\alpha Z) \left[ F_5(\delta) + 2 q F_4(\delta) + q^2 F_3(\delta) \right]~,
\end{equation}
where we have defined $\delta = \eta - q$.

The value of $I$ for beta decay can be written as
\begin{equation}
I_\beta = \theta^{-5} \exp(\upi\alpha Z)
 \int_{\theta}^{q} \frac{\epsilon^2 (\epsilon - q)^2}{1 + \exp[-(\epsilon - \eta)]} d\epsilon~,
\end{equation}
and the value of $J$ for beta decay can be written as
\begin{equation}
J_\beta = \theta^{-6} \exp(\upi\alpha Z)
 \int_{\theta}^{q} \frac{\epsilon^2 (\epsilon - q)^3}{1 + \exp[-(\epsilon - \eta)]} d\epsilon~.
\end{equation}
These integrals can be rewritten (using the substitution
$x = -\epsilon + q$) to be
\begin{equation}
I_\beta = \theta^{-5} \exp(\upi\alpha Z)
 \int_{0}^{q-\theta} \frac{(x - q)^2 x^2}{1 + \exp[x - (q-\eta)]} d\epsilon~,
\end{equation}
and
\begin{equation}
J_\beta = \theta^{-6} \exp(\upi\alpha Z)
 \int_{0}^{q-\theta} \frac{(x - q)^2 x^3}{1 + \exp[x - (q-\eta)]} d\epsilon~.
\end{equation}
We can now make use of the identity
\begin{equation}
  \label{eq:14}
  \int_0^b \frac{x^k}{1 + \exp(x-y)} = F_k(y) - \sum_{j=0}^k \binom{k}{j} b^{k-j} F_j(y - b) ~,
\end{equation}
where we identify $y = q - \eta$ and $b = q-\theta$. The Fermi
integrals in the sum (those with argument $y-b$) will be negligible
because $\theta - \eta \ll -1$ and $F_k(-z) \propto \exp(-z)$.  In
other words, we can extend the upper limit to $\infty$ without
incurring substantial error.  Doing so gives
\begin{equation}
I_\beta = \theta^{-5} \exp(\upi\alpha Z) \left[ F_4(-\delta) - 2 q F_3(-\delta) + q^2 F_2(-\delta) \right]~,
\end{equation}
and
\begin{equation}
J_\beta = \theta^{-6} \exp(\upi\alpha Z) \left[ F_5(-\delta) - 2 q F_4(-\delta) + q^2 F_3(-\delta) \right]~,
\end{equation}
where we have again defined $\delta = \eta - q$.

We are interested in the expression
 \begin{equation}
C = \frac{\varepsilon_{\nu, \mathrm{ec}} \lambda_\beta + \varepsilon_{\nu, \beta} \lambda_\mathrm{ec}}{\lambda_\beta + \lambda_\mathrm{ec}} =  \me c^2 \ln(2) \left(\frac{I_\mathrm{ec} J_\beta + I_\beta J_\mathrm{ec}}{(ft)_{\beta} I_\mathrm{ec} + (ft)_{\mathrm{ec}} I_\beta}\right)~.
\end{equation}
The limit of interest is $q \gg 1$ and $|\delta| < 1$.  Recall that
for $y \ll 1$, $F_k(y) \approx -y \Gamma(k+1)$.  Therefore, after retaining the dominant terms,
\begin{equation}
  \label{eq:c-full}
C = \me c^2 \ln(2) \theta^{-6} q^2 \exp({\upi\alpha Z})\left[\frac{F_2(\delta) F_3(-\delta) + F_2(-\delta) F_3(\delta)}{(ft)_{\beta}  F_2(\delta) + (ft)_{\mathrm{ec}} F_2(-\delta)}\right] ~.
\end{equation}
Evaluating the term in square braces at $\delta = 0$ gives
\begin{equation}
C = \frac{ \me c^2 \ln 2}{(ft)} \theta^{-6} q^2 \exp({\upi\alpha Z})
\left[\frac{7 \upi ^4}{60} \frac{1}{(ft)_{\beta} + (ft)_{\mathrm{ec}} }\right].
\end{equation}
and the peak value of the Urca-process cooling rate
is thus
\begin{equation}
  C_\mathrm{max} = \frac{ 7 \upi ^4 \ln 2}{60}\frac{ \me c^2}{(ft)_{\beta} + (ft)_{\mathrm{ec}} } \left(\frac{\kB T}{\me c^2}\right)^4 \left(\frac{Q}{\me c^2}\right)^2 \exp({\upi\alpha Z}) ~.
\end{equation}
Assuming $(ft)_{\beta} = (ft)_{\mathrm{ec}}$, which is true when the
ground states have the same spins, we can Taylor expand the term in
square braces in equation~\eqref{eq:c-full} to second order to obtain
the dependence of $C$ on $\delta$, the dimensionless energy difference
away from threshold:
\begin{equation}
C \propto \frac{1}{(ft)}
\left[\frac{7 \upi ^4}{120} - \frac{\upi^2 \delta^2}{4}\right].
\label{eq:c-taylor}
\end{equation}
The term in square braces is zero when $\delta = \sqrt{7/30} \upi$,
implying that the characteristic width of the Urca-process cooling
peak is $\approx 3 \delta$, that is when
$\mu \approx |Q| \pm 1.5 \kB T$.

\section{Convergence}
\label{sec:convergence}

In order for the results of our \mesa\ calculations to be meaningful,
we must ensure that the resolution (in both space and time) is
sufficient to resolve the processes of interest.  Once that condition
is achieved, we must also demonstrate that the answer is independent
of the resolution.

The overall spatial and temporal convergence settings used in our
\mesa\ calculations are
\begin{verbatim}
    varcontrol_target = 1e-3
    mesh_delta_coeff = 1.0
\end{verbatim}
Because the weak reactions produce temperature and composition
changes, the default controls typically do an acceptable job of
spatially resolving the cooling and heating regions.  However, the
effective timestep limit in a run with these controls alone is
typically due to the Newton-Raphson solver taking an excessive number
of iterations to converge and \MESA\ limiting the timestep in
response.  It is more satisfying to limit the timestep based on a
physical criteron.  In \SQB, we demonstrated that this value of
\texttt{varcontrol\_target}, along with a timestep criterion based on
changes in central density
\begin{verbatim}
    delta_lgRho_cntr_hard_limit = 3e-3
    delta_lgRho_cntr_limit = 1e-3
\end{verbatim}
gave a converged result.  In this Appendix, we demonstrate that this
is still true when including Urca-process cooling, and we adopt these
as our fiducial resolution controls.

From Section~\ref{sec:urca} and equation~\eqref{eq:c-taylor} above, we know that
the Urca-process cooling occurs over a range corresponding to a   change
in Fermi energy $\Delta \EF \sim \kB T$.  \mesa\ calculates the value
of the quantity $\eta = (\kB T/\EF)^{-1}$ in each cell at each
timestep.  If we ensure the mesh points in our model are selected as
to limit variation of $\Delta \eta$ between adjacent cells and ensure
that our timestep is such that $\delta \eta$ in a given cell between
timesteps is also limited, we will resolve the Urca process.

The scheme by which the spatial resolution in \mesa\ is modified is
described in section 6.5 of \citet{Paxton2011}.  \mesa\ allows the user
to specify other ``mesh functions'' whose cell-to-cell variation will
be reduced below the value of \mdc\ during remeshes.  Therefore, we
define one of the mesh functions to be
$f_1 = \eta / \Delta \eta_{\mathrm{limit}}$.  Then \mesa\ will limit the
change in $\eta$ between adjacent cells $k$ and $k+1$ at timestep $i$,
\begin{equation}
  \Delta \eta = \left|\eta_{k+1}^{i} - \eta_k^{i}\right|~,
\end{equation}
to be less than $\Delta \eta_{\mathrm{limit}}$.

We similarly limit the timestep.  After the solver has taken the
values at timestep $i$ and returned a proposed solution at timestep
$i+1$, we calculate the change in $\eta$ in each cell $k$ and take the
maximum,
\begin{equation}
  \delta \eta = \max\left(\left|\eta_k^{i+1} - \eta_k^{i}\right|\right) ~.
\end{equation}
If $\delta \eta > \delta \eta_{\mathrm{limit}}$, then the proposed
step is rejected and redone with a shorter timestep.  This is similar
to the way variations in the structure variables are limited via
\vt.

\begin{figure}
  \centering
  \includegraphics[width=\columnwidth]{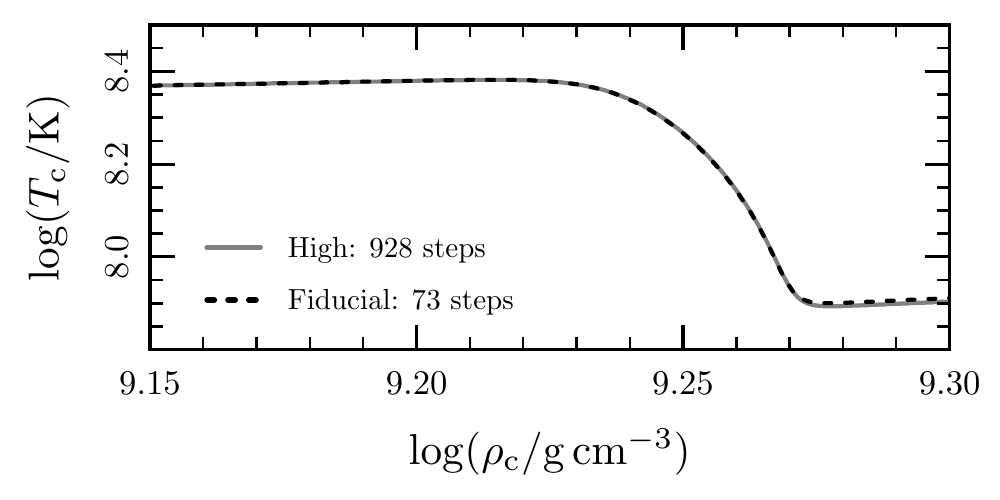}
  \includegraphics[width=\columnwidth]{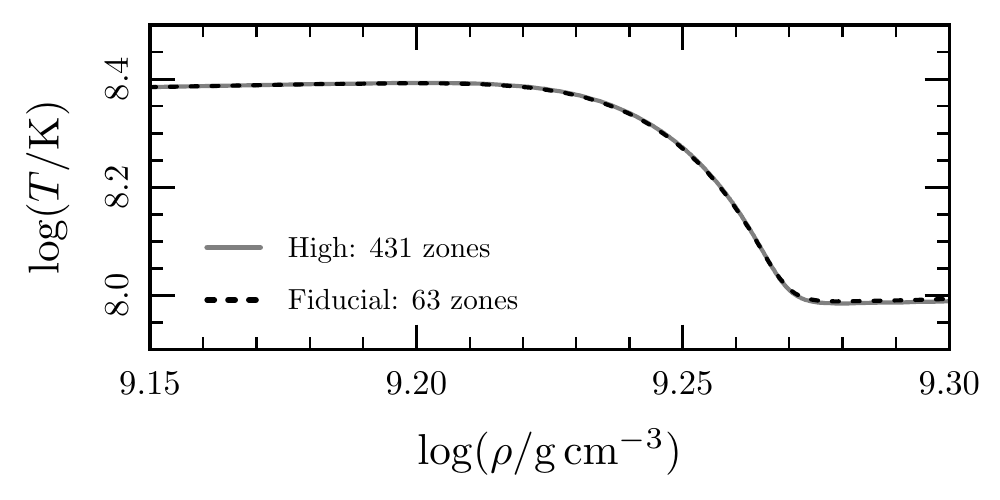}
  \caption{The evolution of a model with $X_\mathrm{Na} = 0.01$ using
    the different resolution controls discussed in the text.  The top
    panel shows the evolution of the central density and temperature.
    The legend shows the number of timesteps used to go from the
    (local) maximum temperature to the (local) minimum temperature.
    The bottom panel shows the density and temperature profile of the
    model when $\logRhoc = 9.4$.  The legend shows the number of
    mesh points covering the region from the (local) maximum
    temperature to the (local) minimum temperature.}
  \label{fig:convergence-toy}
\end{figure}

We vary the spatial and temporal parameters and check that our results
are unaffected.  We use the same model as in Fig.~\ref{fig:schematic},
one composed of \oxygen, \neon, and \sodium\ (with
$X_\mathrm{Na} = 0.01$).  Fig.~\ref{fig:convergence-toy} compares a
run with the limits $\Delta \eta_{\mathrm{limit}} = 1$ and
$\delta \eta_{\mathrm{limit}} = 1$ to our fiducial resolution
controls, which limit the change in central density as in \SQB.  The
results of the fiducial and high resolution cases are nearly
indistinguishable in the quantities of interest, demonstrating that
our results are converged.

\section{Toy Model of Runaway}
\label{sec:toy-runaway}

In order to gain insight into the behavior observed in our \MESA\
calculations, we use a toy model of a thermal runaway process.  We
solve a reaction-diffusion equation
\begin{equation}
  \frac{\partial T}{\partial t} - K \nabla^2 T = -q \frac{dY}{dt}~,
  \label{eq:toy-model}
\end{equation}
where $T$ represents the temperature, $K$ the thermal conductivity,
and $Y$ the abundance.\footnote{To solve this PDE, we use \texttt{dedalus} \citep{Burns2018}; \url{http://dedalus-project.org}}
We non-dimensionalize $T$ and $Y$ by their
initial values and begin from uniform initial conditions, so
$T(r, t=0) = 1$ and $Y(r, t=0) = 1$.  The radial extent sets our
length scale, so we solve on the domain $r \in [r_\epsilon,1]$, where
the choice of $r_\epsilon = 10^{-4}$ avoids difficulties associated
with the coordinate singularity at $r=0$.  The value of $q$ encodes
temperature change due to energy release from the reaction consuming
$Y$; in the absence of diffusive transport, a parcel would reach a
temperature of $T = 1+q$ once $Y=0$.

We choose the reaction rate for $Y$ to have the form of a
sub-threshold electron capture rate
\begin{equation}
  \frac{dY}{dt} = Y T^3 \exp\left(\frac{\Delta(r)}{T} - \Delta_0\right)~,
  \label{eq:toy-rate}
\end{equation}
where physically $\Delta$ represents how close the chemical potential
is to the threshold chemical potential in units of $\kB T$.  The
inclusion of $\Delta_0 \equiv \Delta(0)$ ensures that at $r=0$ and
$t=0$ we have $\frac{dY}{dt}= 1$ (i.e. we non-dimensionalize using the
initial reaction time-scale in the centre).

In our stellar models, where the pressure is dominated by degenerate,
relativistic electrons
$P \approx P(\rho) \propto \rho^{4/3} \propto \mu^4$ (where $\mu$ is
the electron chemical potential).  Hydrostatic equilibrium implies
that $\lim_{r \to 0} \frac{dP}{dr} = 0$.  Therefore, we assume
\begin{equation}
  \Delta(r) = \Delta_0 - \Delta_2 r^2
\end{equation}
This spatial variation in $\Delta$ is what will lead to the thermal
runaway.  At $t=0$, the reaction rate is a factor of $e$ lower at
$x = \sqrt{1/\Delta_2}$; this sets the initial length scale of the
runaway.  We chose the following fiducial parameters
\begin{align}
  q &= 3 ~,\\
  \Delta_0 &= -7~,\\
  \Delta_2 &= 10~,
\end{align}
which are in rough quantitative agreement with the physical parameters
whose effects they represent.

We want to use these models to inform our understanding of the
convective instability of our \MESA\ models.  Because the toy model
does not include density or gravity, one cannot directly assess its
stability. However, we understand that in the stellar models the
stability is determined largely by the temperature and composition
gradients.  The Ledoux criterion for convective instability is
\mbox{$B < \delta_\nabla$}.  The temperature gradient sets
\begin{equation}
  \delta_\nabla \equiv \gradT - \gradad \approx -\frac{H_P}{T} \frac{dT}{dr} \propto -\frac{1}{T} \frac{dT}{dr}~,
  \label{eq:superad-proxy}
\end{equation}
where we have assumed $\gradT \gg \gradad$.  The composition gradient
sets
\begin{equation}
  B \equiv  -\frac{1}{\chi_T}
      \left(\frac{\partial \ln P}{\partial \ln \Ye}\right)_{\rho,T}
      \frac{d \ln \Ye}{d \ln P} \approx \frac{\bar{Z} \EF}{3 \kB T} \frac{H_P}{\Ye} \frac{d\Ye}{dr} \propto \frac{1}{T} \frac{dY}{dr}~,
  \label{eq:B-proxy}
\end{equation}
where we have assumed that the total change in $\Ye$ due to the change
in $Y$ is small compared to $\Ye$ itself.  We will refer to the
expressions to the right of the proportionality signs in
equations~\eqref{eq:superad-proxy} and \eqref{eq:B-proxy} as our
``proxies'' for $\delta_\nabla$ and $B$.  These ``proxies'' allow us
to understand how the gradients evolve in relation to one another.
Our primary interest is the behavior of the centre, so we measure
these values at $r = 10\,r_\epsilon$.

\subsection{No diffusion ($K=0$)}

\begin{figure}
  \centering
  \includegraphics[width=\columnwidth]{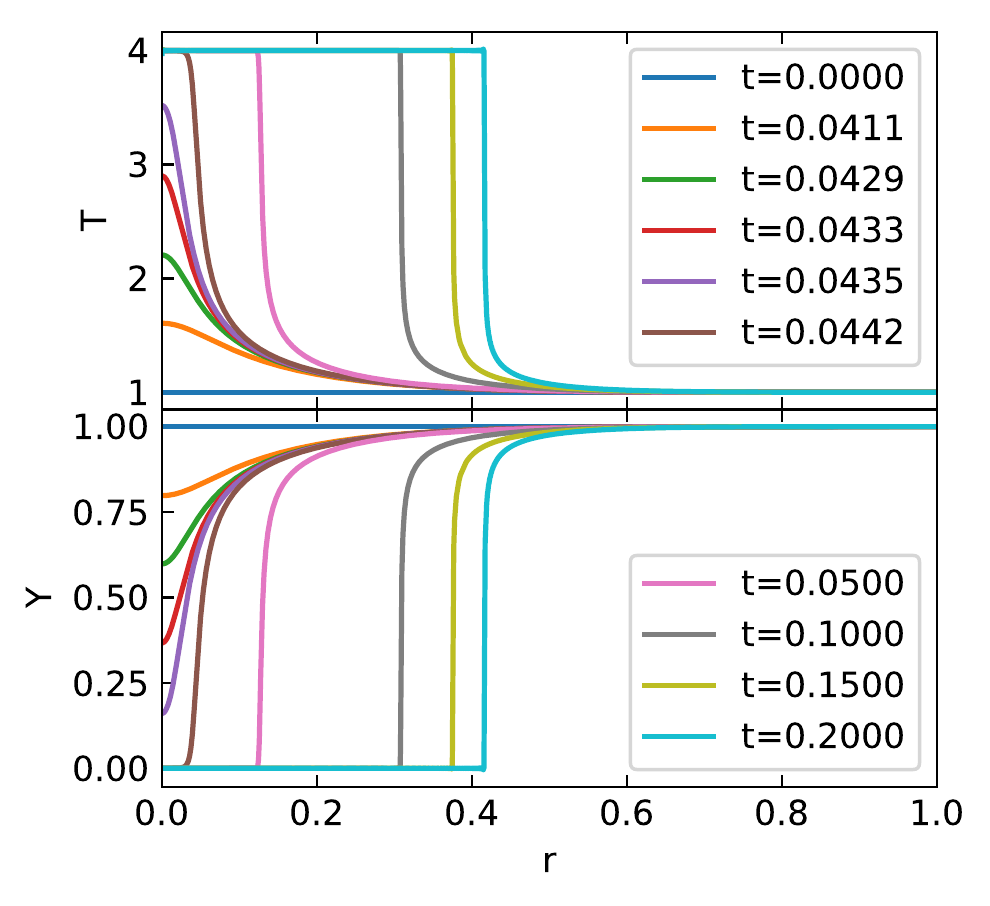}
  \caption{Runaway in the absence of thermal conduction $(K=0)$.  The
    top panel shows the temperature and bottom panel shows the
    composition.  It takes longer for the runaway to complete at
    larger radii (due to the lower chemical potential), so the region
    of completion moves outward with time. }
  \label{fig:nodiffusion}
\end{figure}

First, we study this problem in the absence of diffusion.  In this
case, parcels at different $r$ evolve independently.  The temperature
of a parcel is therefore given by $T = 1 + q (1-Y)$.  Formally, it
takes infinite time to reach $Y=0$; however, in practice this poses no
problem, as arbitrarily small values of $Y$ are reached in finite
time.  The time-scale for the central parcel to reach $Y\approx 0$
$(Y = 10^{-4})$ is $t_{\mathrm{runaway}} \approx
0.044$. Fig.~\ref{fig:nodiffusion} shows the $T$ and $Y$ profiles for
a range of times.

As discussed previously, the runaway is seeded on a length scale
$l_{\mathrm{runaway}} = \sqrt{1/\Delta_2} \approx 0.3$.  In the early
phase of the runaway the length scale shrinks.  As $Y$ is depleted,
the reaction rate eventually ceases increasing and begins to decrease.
This happens first to parcels in the centre and so the length scale
begins to increase as off-centre parcels begin to catch up.  This
implies there is some minimum length scale, and for the fiducial
parameters this is $l_{\mathrm{min}} = 0.032$.

\begin{figure}
  \centering
  \includegraphics[width=\columnwidth]{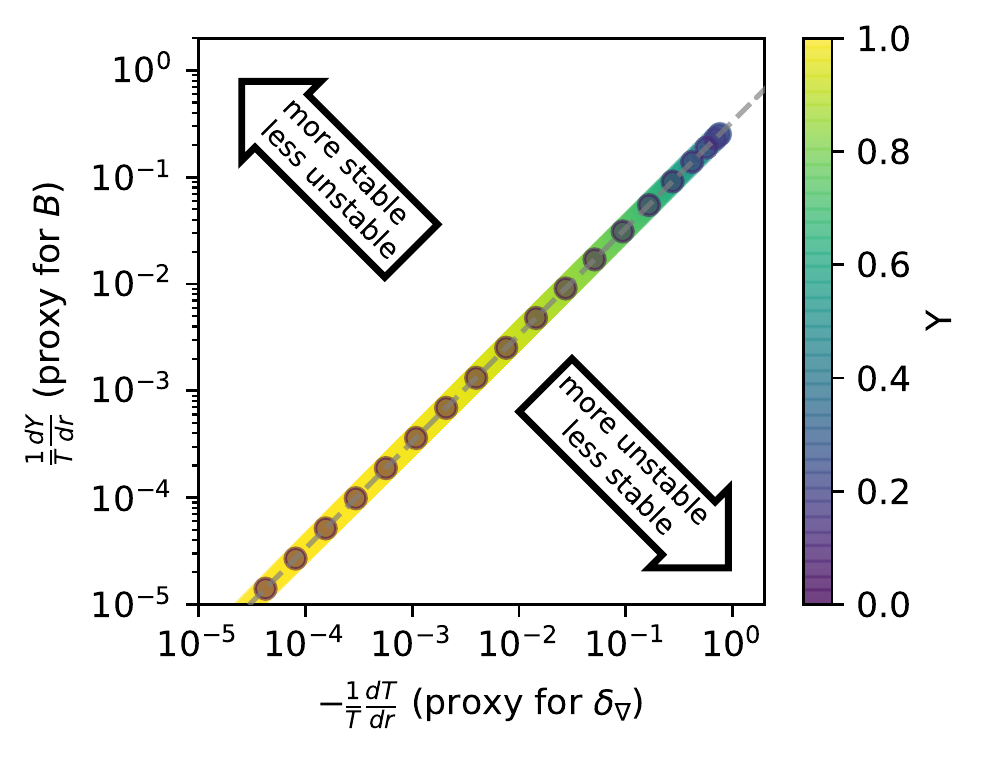}
  \caption{Stability in the absence of thermal conduction ($K = 0$).  The points show the values in our
    calculation, regularly spaced in time.  The color gives the value
    of $Y$, as indicated by the color bar.  The model begins in the
    lower left (yellow, $Y=1$) and moves up and to the right as $Y$
    decreases.  As $Y$ decreases further (dark blue/purple,
    $Y \la 0.1$), the model reverses and moves back down again towards
    the lower left.  The apparent transition of the points from
    continuous to discrete as $Y$ decreases is a consequence of the
    decrease in the time-scale as the runaway proceeds.  The grey
    dashed line shows the analytically expected constant ratio of the
    T and Y gradients.  The arrows indicate the direction in which
    stability changes; as discussed in the text, we can make only
    relative statements about stability.  If initially stable, this
    remains stable.}
  \label{fig:stability-nodiffusion}
\end{figure}

Fig.~\ref{fig:stability-nodiffusion} shows our proxies for
$\delta_\nabla$ and $B$ as a function of $Y$.  Note that this figure
and the others like it are log-log plots.  Thus a true plot of
$\delta_\nabla$ vs. $B$ would have the same shape, as the constants of
proportionality act as translations.  Since $T$ is a linear function
of $Y$, the gradients have a constant ratio, $(dT/dr)/(dY/dr) = -q$.
This relationship is shown as a grey dashed line and it is clear that
it holds at all times during the evolution.

\subsection{Infinitesimal diffusion ($K=\epsilon$)}

\begin{figure}
  \centering
  \includegraphics[width=\columnwidth]{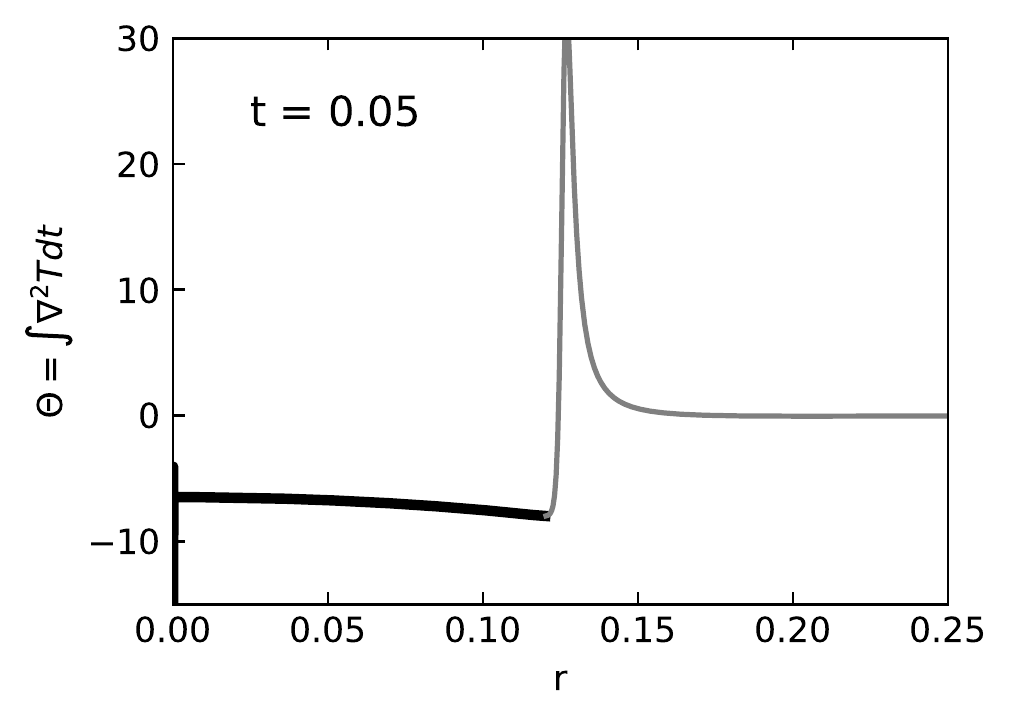}
  \caption{The thick black portion of the line marks where
    $Y < 10^{-4}$.  In these regions the runaway is finished and
    $\Theta$ will no longer evolve with time.}
  \label{fig:theta}
\end{figure}

In the presence of an infinitesimally small diffusion coefficient, the
temperature evolution of the runaway would remain unchanged.
Therefore, we can use the results of the $K=0$ calculation to evaluate
the effect of small diffusion coefficients.  The sharp temperature
gradient at the transition edge leads to heating of the fluid element
in advance of the transition, followed by later cooling as it gives
the heat back.  The change in temperature due to conduction at a
location $r$ between the start of the calculation and a time $t$ is
given by $K\Theta(r,t)$, where
\begin{equation}
  \Theta(r,t) = \int_0^t \nabla^2 T(r,t') dt' ~.
  \label{eq:theta-defn}
\end{equation}
For values of $t$ in excess of the time it takes the runaway to
complete at a location $r$, the value of $\Theta$ will no longer
evolve.  Fig.~\ref{fig:theta} shows $\Theta$ at $t = 0.05$.  The
region where the runaway is finished is marked by the bold black line;
regions outside of this location are still ``active'' in terms of heat
transfer.  Note that in the toy model $r$ is a Lagrangian coordinate.

Fig.~\ref{fig:theta} shows that $\Theta$ is
negative near the centre ($r \la 0.1$), indicating that heat is conducted out of the core.  More
importantly, it shows that $\Theta$ decreases with increasing $r$.
This indicates that conduction will cause a residual temperature
gradient after the runaway.  Taking the time integral (as in
equation~\ref{eq:theta-defn}) of all terms in
equation~\eqref{eq:toy-model} gives
\begin{equation}
  T(r,t) - 1 - K \Theta(r,t) = -q \left(1-Y(r,t)\right)
\end{equation}
When the runaway has finished, $Y \approx 0$, and this implies that
$\frac{dT}{dr} = K \frac{d\Theta}{dr}$ in these regions.  Thus, at the
end of the runaway, when the composition gradient has vanished, a
residual temperature gradient can remain.  Fig.~\ref{fig:theta}
indicates that this temperature gradient is radially decreasing and
thus has the potential to lead to the onset of convective instability.

\subsection{Finite diffusion ($K >0$)}
\label{sec:finite-diffusion}

The approximation that thermal diffusion does not affect the runaway
must be reasonable only for $K$ less than some value
$K_{\mathrm{crit}}$.  We now estimate this critical value in two ways.
From Fig.~\ref{fig:theta} we can estimate that the size of the
conductive temperature perturbation is $\approx 6 K$.  Changing the
rate given in equation~\eqref{eq:toy-rate} by $e$ requires a
temperature change $\approx T^2/ \Delta_0$.  For $T \approx 2$ (the
geometric mean of the initial and final temperature) this is
$\approx 0.6$.  Equating these temperature changes suggests a value
$K_{\mathrm{crit}} \sim 0.1$.  Physically, the characteristic length
and time-scales associated with the runaway also give a estimate
\begin{equation}
  K_{\mathrm{crit}} \sim \frac{l_{\mathrm{min}}^2}{t_{\mathrm{runaway}}} \approx 0.02 ~.
\end{equation}
for when conduction will modify the runaway.  These estimates agree
and so to demonstrate the effects of conduction we solve our toy
problem for $K = 0.01$ and $K = 0.1$.

\begin{figure}
  \centering
  \includegraphics[width=\columnwidth]{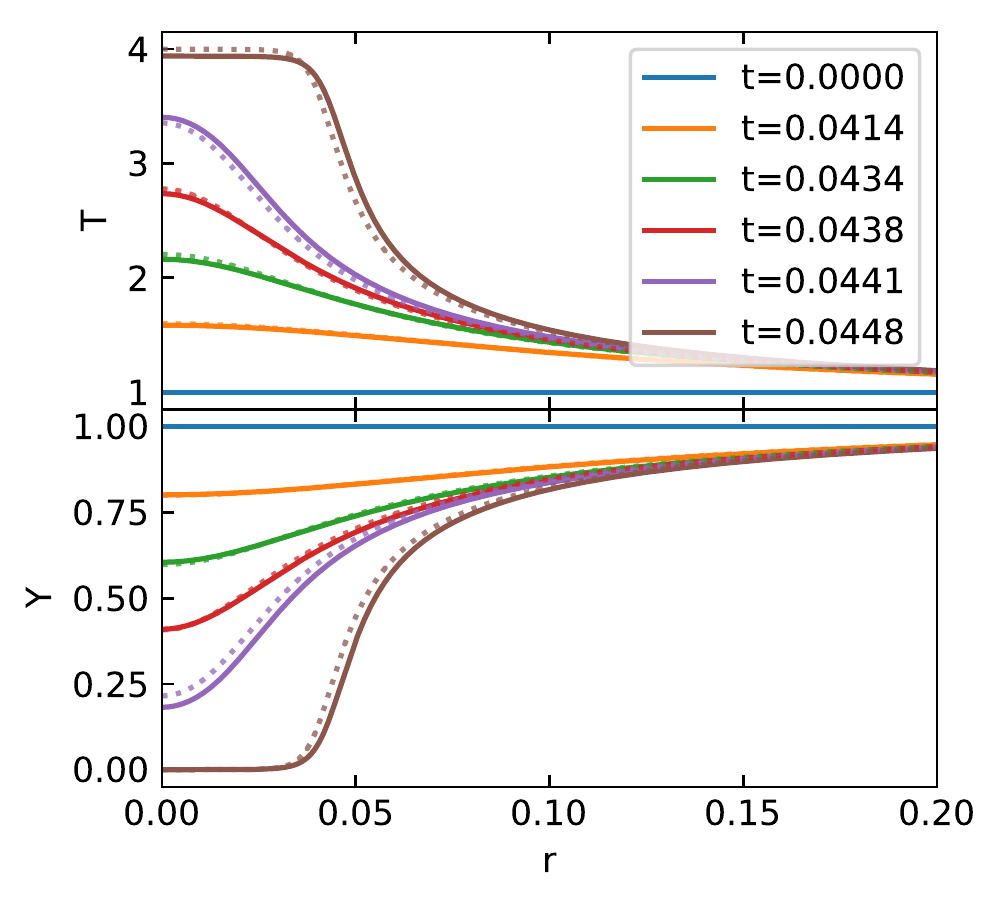}
  \caption{Runaway including thermal conduction for
    $K < K_{\mathrm{crit}}$ $(K=0.01)$.  Same as
    Fig.~\ref{fig:nodiffusion}, but zoomed in on the central region.
    The dotted lines show the profiles from the model without
    conduction $(K=0)$ from times with approximately matching central
    values of $Y$.  Conduction has not significantly modified the
    runaway.}
  \label{fig:lodiff}
\end{figure}

\begin{figure}
  \centering
  \includegraphics[width=\columnwidth]{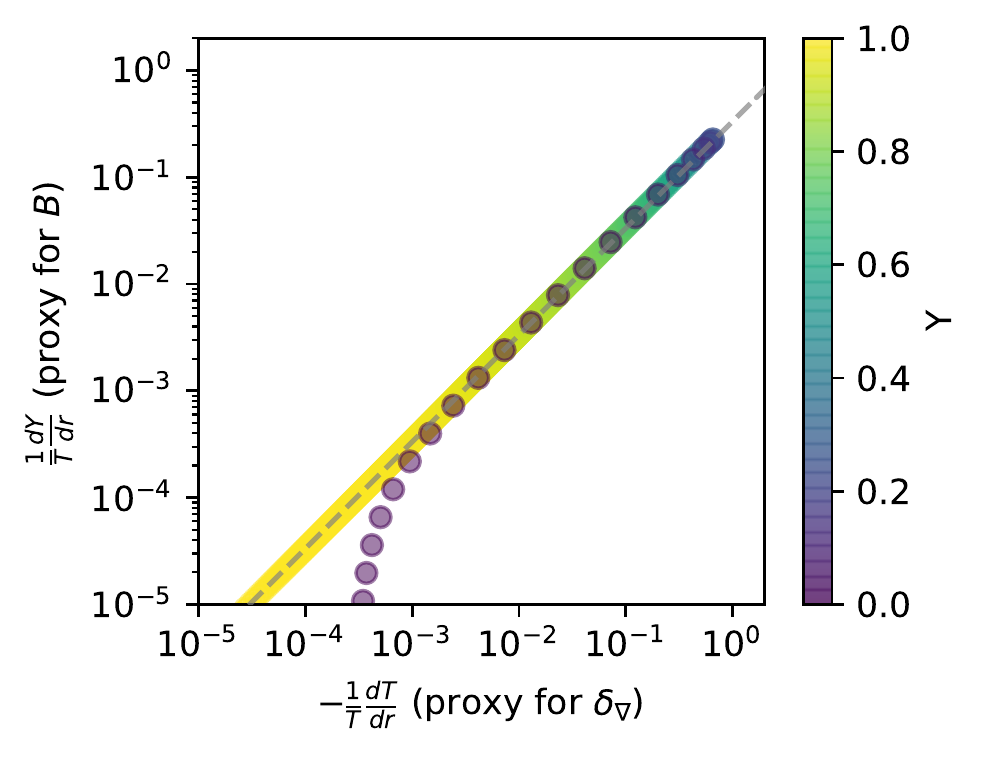}
  \caption{Stability including thermal conduction for $K < K_{\mathrm{crit}}$ $(K=0.01)$,
    visualized as in Fig.~\ref{fig:stability-nodiffusion}.  Note that
    as the Y gradient vanishes, a T gradient remains.  Thus even if
    initially stable, this can evolve to become unstable.}
  \label{fig:stability-lodiff}
\end{figure}

Fig.~\ref{fig:lodiff} shows the $T$ and $Y$ profiles for $K=0.01$.
Conduction has not significantly modified the runaway.
Fig.~\ref{fig:stability-lodiff} shows our stability diagnostic plot
for this case.  The solution evolves with a constant ratio of
$|dT/dr|/|dY/dr|$ (same trajectory as $K=0$), until $Y \ll 1$ at which
point this ratio begins to increase.  This is evolving in the
direction of instability.

\begin{figure}
  \centering
  \includegraphics[width=\columnwidth]{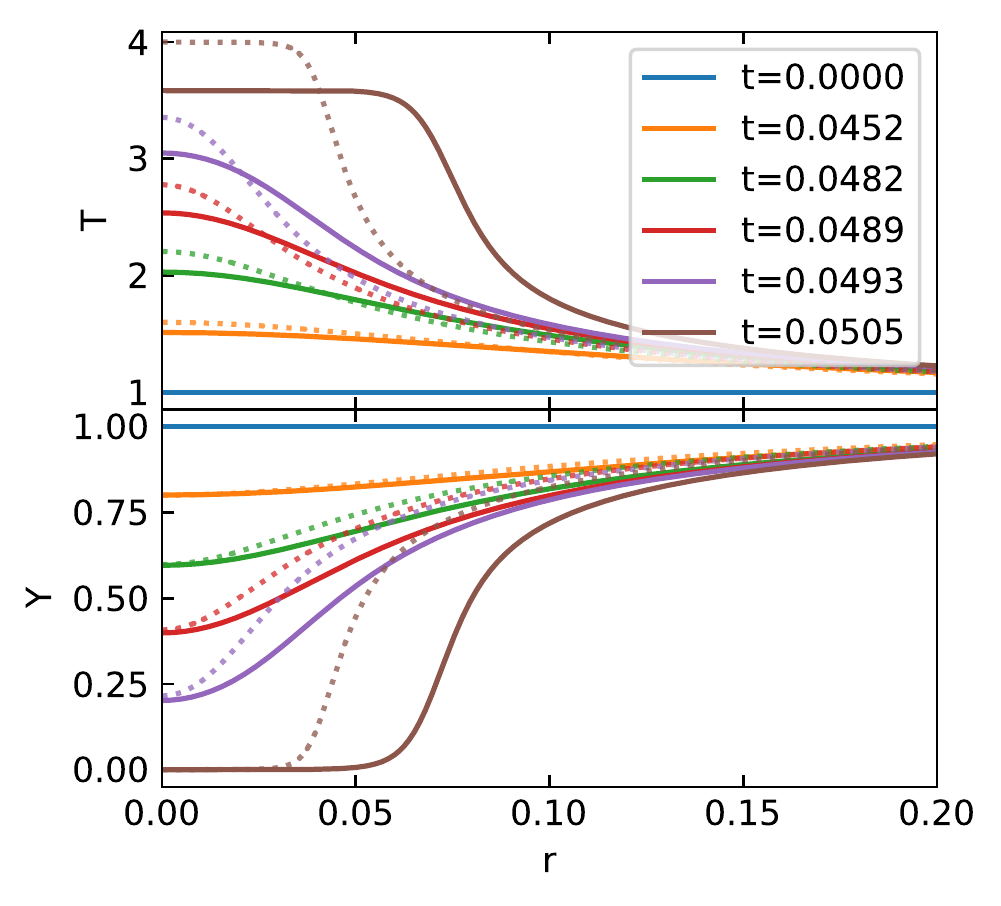}
  \caption{Runaway including thermal conduction for
    $K \gtrsim K_{\mathrm{crit}}$ $(K=0.1)$.  Same as
    Fig.~\ref{fig:nodiffusion}, but zoomed in on the central region.
    The dotted lines show the profiles from the model without
    conduction $(K=0)$ from times with approximately matching central
    values of $Y$.  Conduction has significantly modified the
    runaway.}
  \label{fig:hidiff}
\end{figure}

\begin{figure}
  \centering
  \includegraphics[width=\columnwidth]{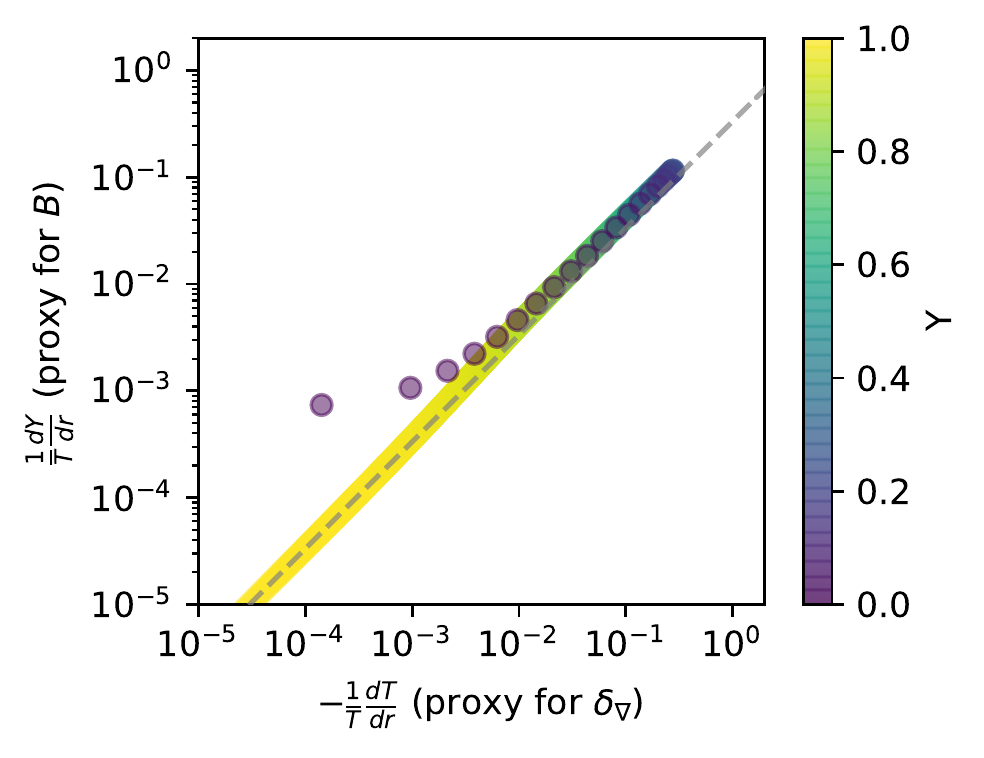}
  \caption{Stability including thermal conduction for $K \gtrsim K_{\mathrm{crit}}$ $(K=0.1)$,
    visualized as in Fig.~\ref{fig:stability-nodiffusion}. The T
    gradient is always less than it would be in the absence of
    conduction.  Thus if initially stable, this remains stable.}
  \label{fig:stability-hidiff}
\end{figure}

Fig.~\ref{fig:hidiff} shows the $T$ and $Y$ profiles for $K=0.1$.
Conduction has significantly modified the runaway.
Fig.~\ref{fig:stability-hidiff} shows our stability diagnostic plot
for this case.  The solution departs from the $K=0$ trajectory even
for $Y \approx 1$, where ratio of $|dT/dr|/|dY/dr|$ decreases,
indicating that conduction makes things more stable.

This demonstrates that a thermal runaway driven by sub-threshold
electron captures in which thermal conduction operates can lead to
convective instability at the centre of a star.

\subsection{Connection to \MESA\ Models}

In order to complete the connection with our \MESA\ models, we
estimate the value of $K$.  Since the toy equations are not the same
as the equations solved by \MESA, this estimate is done at the order
of magnitude level.  This approximate value $K_{\MESA}$ is the
appropriately non-dimensionalized version of the thermal diffusivity
in the star.

In the dimensionless units associated with the toy problem, we
observed the runaway had a minimum length scale of 0.03 and a time
scale of 0.05.  In the \MESA\ calculation shown in
Fig.~\ref{fig:runaway}, the runaway has a minimum length scale of
$\unit[3\times10^5]{cm}$ and a time-scale of $\unit[50]{yr}$.  That
suggests that the time and length scales with which one should
non-dimensionalize are $\unit[10^7]{cm}$ and $\unit[10^3]{yr}$

The thermal diffusivity at the relevant conditions is
$\approx \unit[60]{cm^2\,s^{-1}}$.  (This is the value returned by the
\MESA\ \texttt{kap} module, which uses the results from
\citet{Cassisi2007}, for $\logRho \approx 9.6$ and $\logT \approx 8.4$
with a 50/50 oxygen-neon mixture.)  So we have
\begin{equation}
  K_{\MESA} \sim \frac{D_{\rm th}}{\rm [L]^2/[T]} \sim 
  \frac{\unit[60]{cm^2\,s^{-1}}}{\unit[3\times10^3]{cm^2\,s^{-1}}} \sim 0.02 ~.
\end{equation}
The range of estimates for $K_{\rm crit}$ found in Section~\ref{sec:finite-diffusion} was 0.02 - 0.1.
This indicates that the \MESA\ models are in the regime of finite
conductivity, but with $K_{\MESA} \la K_{\rm crit}$.  Therefore, this
toy calculation explains the formation of a central convection zone in
our \MESA\ models (see Fig.~\ref{fig:unstable_regions}).

\section{Comparison with models calculated using tabulated rates}
\label{sec:suzuki-tables}

Recently, \citet{Suzuki2016} computed weak reaction rates for the
sd-shell nuclei with mass number A=17-28 using the USDB Hamiltonian.
They include Coulomb effects and take into account experimentally
measured energies and Gamow-Teller transition strengths.  These rates
are tabulated on a finely-spaced grid of density and temperature.
The primary scientific motivation for these new rate tabulations is the evolution of the degenerate oxygen-neon cores that develop in stars with initial masses $\approx \unit[8-10]{\Msun}$.

\begin{figure}
  \centering
  \includegraphics[width=\columnwidth]{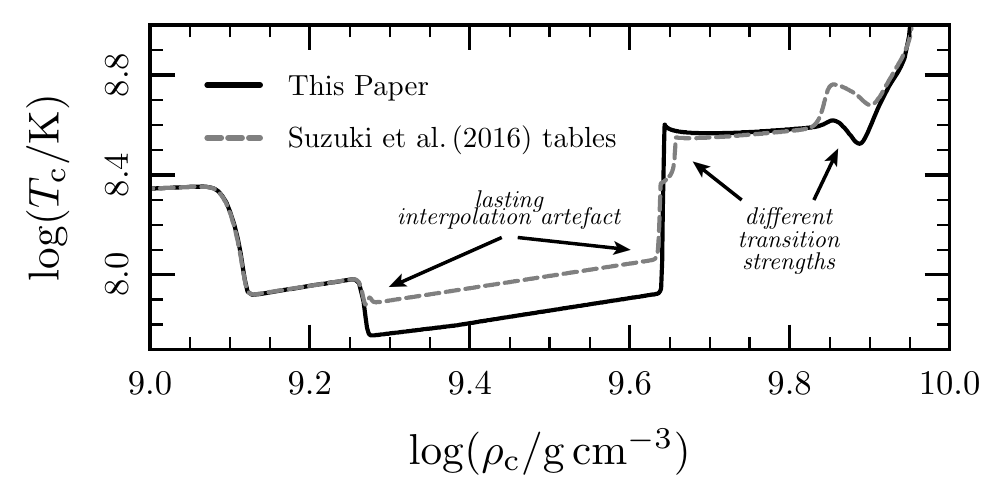}
  \caption{Comparison of the fiducial model presented in this paper
    with a model evolved using the tabulated weak reaction rates from
    \citet{Suzuki2016}.  The models generally agree well.  We describe
    the origin of the indicated differences in the text.}
  \label{fig:suzuki}
\end{figure}

We incorporated these rate tables into \MESA\ and used them in place
of the on-the-fly rates (described in Section~\ref{sec:mesa-options})
to evolve an otherwise identical version of the fiducial model
presented in this paper.  Fig.~\ref{fig:suzuki} compares the central
evolution of a model calculated using these tables with our fiducial
case.  Overall, the agreement is good and there is virtually no
variation in the density at oxygen ignition.  However, there are small
quantitative differences.

The models agree almost perfectly throughout the
\magnesium[25]-\sodium[25] Urca cooling (around
$\logRhoc \approx 9.1$).  This indicates that our on-the-fly rates
agree extremely well with the tabulated rates.  Differences in the
Coulomb corrections would manifest as a shift in density; differences
in transition strengths would appear as shifts in temperature.  No such differences are seen.

The models begin to disagree near the end of the \sodium[23]-\neon[23]
Urca cooling (around $\logRhoc \approx 9.27$).  This reflects the fact
that the model has become so cold that even the finely-sampled table
of \citet{Suzuki2016} is suffering from the interpolation issues
discussed by \citet{Fuller1985} and \citet{Toki2013}.  The
\citet{Suzuki2016} tables are constructed such that these issues
\textit{do not} arise in stars that develop degenerate ONe cores,
where the temperatures typically remain $\ga \unit[3\times10^8]{K}$.
However, in our more demanding application, we reach temperatures
below $\unit[10^8]{K}$.  The extent of the Urca cooling region in
density is $\Delta \ln \rho \approx 9 (\kB T/\EF)$, which is
$\approx 0.01$ at these conditions.  This is now below the table
spacing in this region, which is $\Delta \ln \rho \approx 0.046$.  The
on-the-fly rates avoid interpolation issues and so our models are more
accurate in this regime.

The higher temperature at the end of the $A=23$ Urca cooling leads to
the onset of electron captures on \magnesium[24] at a slightly lower
density.  Subsequently, the differences in the two models are
primarily due to differences in the assumed strength of the non-unique
second forbidden transitions (see Section~\ref{sec:forbidden}).  The
\sodium[24]-\neon[24] non-unique second forbidden transition is not
included in \citet{Suzuki2016}; the \neon[20]-\fluorine[20] non-unique
second forbidden transition is included at the experimental upper
limit.  This disagreement is the result of physical ignorance, and so
we would not favor one result over the other.  One of the motivations
for using the on-the-fly rates is the ease with which one can vary
experimentally-uncertain transition strengths and thus characterize
their effects.


\bsp	
\label{lastpage}
\end{document}